\documentclass{llncs}
\usepackage{amssymb}
\usepackage{fancyvrb}
\usepackage{rotating}
\usepackage{paralist}
\usepackage{colortbl}
\usepackage{algorithm}
\usepackage{algorithmic}
\usepackage{epsfig}
\usepackage{graphicx}
\usepackage{color}
\usepackage{multirow}
\usepackage{tabularx}
\usepackage{epstopdf}
\usepackage{url}

\usepackage{amsmath}
\usepackage[compatibility=false]{caption}
\DeclareCaptionType{copyrightbox}
\usepackage{subcaption}
\usepackage{cleveref}
\usepackage{float}

\usepackage{adjustbox}

\DeclareMathOperator{\Bernoulli}{Bernoulli}

\crefname{algorithm}{alg.}{algs.}
\Crefname{algorithm}{Algorithm}{Algorithms}
\crefname{thm}{thm.}{thms.}
\Crefname{thm}{Theorem}{Theorems}

\newcommand{\para}[1]{{\vspace{4pt} \bf \noindent #1 \hspace{10pt}}}
\newcommand{\fixme}[1]{#1}
\newcommand{\cutme}[1]{{\color{blue} }}

\newcolumntype{H}{>{\columncolor{black}\color{white}}c}

\graphicspath{{./figures/}}

\begin{document}

\title{Profiling User Activities \\With Minimal Traffic Traces}


\author{Tiep Mai \and Deepak Ajwani \and Alessandra  Sala}
\institute{Bell Laboratories, Ireland\\
\url{{tiep.mai,deepak.ajwani,alessandra.sala}@alcatel-lucent.com}}

\maketitle

\begin{abstract}
Understanding user behavior is essential to personalize and enrich a
user's online experience. While there are significant benefits to be
accrued from the pursuit of personalized services based on a
fine-grained behavioral analysis, care must be taken to address user
privacy concerns. In
this paper, we consider the use of web traces with 
truncated URLs -- each URL is trimmed to only contain the web
domain -- for this purpose. While such truncation removes the
fine-grained sensitive information, it also strips the data of many 
features that are crucial to the profiling of user activity. 
We show how to
overcome the severe handicap of lack of crucial features for the
purpose of filtering out the URLs representing a user activity from the noisy network traffic trace (including
advertisement, spam, analytics, webscripts) with high accuracy. This activity profiling with truncated URLs enables the network operators to
provide personalized services while mitigating privacy concerns by
storing and sharing only truncated traffic traces.


In order to offset the accuracy loss due to
truncation, our statistical methodology leverages specialized features extracted from a group of consecutive URLs that represent a
\textit{micro user action} like web click, chat reply, etc., which we call bursts. These bursts, in
turn, are detected by a novel algorithm which is 
based on our observed characteristics of the inter-arrival time of
HTTP records. We present an extensive experimental evaluation on a
real dataset of mobile web traces, consisting of more than 130 million
records, representing the browsing activities of 10,000 users over a
period of 30 days. Our results show that the proposed methodology
achieves around 90\% accuracy in segregating URLs representing user
activities from
non-representative URLs. 
\end{abstract}

\section{Introduction}
\label{sec:introduction}
Behavioral analysis of mobile users based on their web activities has the
potential to transform their online experience. It enables service
providers to personalize their deliverable, specialize their content,
customize recommendations and target advertisements based on user
context. For the network operators, it opens up the possibility of 
provisioning their resources and dynamically managing their network
infrastructure (particularly, with the realization of network function
virtualization) to effectively serve the varying user and content
demand in order to deliver advanced quality-of-service experience.

However, behavioral analysis also raises serious concerns about user
privacy. Users are uncomfortable if personalization is taken too far. In the wider 
philosophical debate between personalized services based on user
behavior analysis and preserving the user privacy, there is a need to
find a middle ground that will allow for potential benefits of
personalized services and still safeguard the fine-grained sensitive
user information.\footnote{ Specific search queries, personal
entertainment preferences, purchased products,
location etc. are generally considered highly sensitive
user information.} 

Ideally, the data set
for such analyses should be stripped of all sensitive user
information, while still allowing for inference of medium-grained user
activity. This is becoming even more important with the 
tightening privacy legislations in various
countries~\cite{hamburgGoogle,EUPrivacyReg,EUPrivacyReg2,bbcPrivacyNews}, increasing regulation
(e.g.,~\cite{FacebookPrivacy2}) and heavy penalties for data breaches which has made network operators as
well as service providers (e.g.,~\cite{FacebookPrivacy1}) more careful about the
data sets they collect, store and share. The operators would like to
store the minimal amount of data to still be able to perform complex
analytics, raising the important question of determining the \textit{thin
boundary} between the required data for necessary analytics and the
data that can enable mining of highly sensitive fine-grained user
traits. In this context, we consider the usage of truncated URLs,
wherein each URL is trimmed to only contain the web
domain. For instance, the
HTTP URL \textit{finance.yahoo.com/q?s=BAC} is truncated to
\textit{finance.yahoo.com} (to hide the fact that the user had queried
for Bank of America Corp. stock price), the URL
\textit{https://www.google.com/\#q=postnatal+depression} is truncated to
\textit{www.google.com} (to avoid leaking the sensitive health query
of the user) and the URL
\textit{www.amazon.com/\allowbreak Dell-Inspiron-i15R-15-6-inch-Laptop/dp/\allowbreak B009US2BKA}
is truncated to \textit{www.amazon.com} (to avoid leaking the
searched or purchased product). Already, many network operators only
share the truncated URL data-sets with third-party analysts,
owing to privacy considerations. For the non-HTTP traces (e.g., HTTPS
encapsulated in IP packets), even the network operators, themselves, have limited
information available. While a reverseDNS service can be used to extract the
URL from the IP address, it does not recovers the content type or the
query parameters. Thus, it is important to explore whether high accuracy can still be obtained in profiling user activities if
an analyst is restricted to only using truncated-URL web trace. In
this paper, we investigate this issue. 

Specifically, we focus on the task of identifying URLs that are representative of user
activities, which is often an important step in profiling user
activities. We note that the remaining task of mapping the
representative URLs to activity categories (and creating user
profiles) can be done using either
manual labeling of interesting categories or in an automated way by
using external databases or web analytics services (e.g.,
Alexa~\cite{alexa}). 

The key challenge in filtering out the representative URLs from noisy
truncated traffic trace is that a truncated trace lacks many crucial
features for such a filtering. These include the file name suffix (e.g., .jpg, .mp3, .mpg etc.) that is usually a good indicator
of the content type as well as number, type and values of parameters
in the URL strings. Nonetheless, we show that even with the truncated URLs, we can
achieve a highly accurate automated classification of web-domains into
those that represent the user activity and those that don't. The key
insight that we bring in this paper is that a user's traffic trace is
composed of many data bursts. A burst usually corresponds to a \textit{micro
user action} like a web click, chat reply, etc. and is typically
associated with a unique activity. We show that novel features related
to burst measurements, such as positioning
of a URL in a burst, the number of URLs in the burst containing the
web-domain etc., can improve the accuracy of filtering the noise
(unintentional traffic such as spam, analytics, advertisements as well as other
non-representative traffic such as images, multimedia, scripts) out
of the traffic trace, by around 20\%, offsetting
the loss due to URL truncation.


\fixme{

To achieve this result, we need to decompose a traffic trace of a user into its constituent
data bursts. The problem here is that there is a significant variation in
the traffic pattern across different users, at different timestamps
and different activities. Even the distribution shape of the
inter-arrival time of HTTP records differs significantly from one user
to another. We resolve this problem by proposing a novel burst
decomposition algorithm that adapts itself to any distribution shape, 
rather than relying on specific distributions.

We provide an extensive experimental evaluation over more
than 130 million HTTP records generated from 10,000 users
over a period of 30 days. The experimental analysis demonstrates that
our methodology provides high accuracy (around 90\%), in segregating
representative URLs from non-representative URLs. 

Our approach, thus, enables the
network operators to personalize services without risking the leakage
of more sensitive user data (as the sensitive information need not be
stored or shared). Specifically, it enables many medium-grained 
personalization applications, including, but not restricted to, product
recommendation and targeted advertisement. For instance, knowing when their users
read, shop, browse and play games, enables telecom operators to create better
pricing schemes that are personalized and targetted for different
users and demographics. Such profiling of user
activity also opens up many avenues for network optimization to
service providers. For instance, system resources can be better allocated to match
the data access rate and desired delay time for gaming activities at
specific time in the day and better caching strategies can be designed. }

\textbf{Outline} In Section \ref{sec:variation}, we show that there is a
considerable variation in the user activity, that necessitates the
data-dependent feature extraction and complex statistical models to
deal with this problem. Section \ref{sec:methodology} presents an overview of our
methodology.  In Section \ref{sec:inter-arrival}, we argue that there
is a considerable variation in the distribution shapes of the
inter-arrival time of HTTP records and thus, the burst decomposition
techniques that rely
on specific distribution shapes do not work well across the entire
user spectrum. In Section \ref{sec:burstdetection}, we show how we can
remedy the situation by using a threshold on the inter-arrival time of HTTP
records, that adapts to the distribution profile of each
user. Section \ref{sec:domainclassification} presents the results of
domain classification using the features extracted from burst
measurements. Section \ref{sec:relatedwork} presents an overview of some related work.

\section{Variation in User Activity}
\label{sec:variation}

The main goal of our investigation is to develop an automated
procedure to filter out representative URLs from the noisy trace of
truncated HTTP records. In this section, we describe our dataset. We
show that in this dataset, there is a significant variation between
different users in terms of non-representative traffic, user
activities, number of HTTP records etc. In the next section, we
propose a novel methodology that employs robust algorithms for
extracting user-dependent features to overcome this high variation 
in user activity.

\para{Dataset.} Our dataset consists of more than 130 million 
web-logs generated from randomly selected 10,000
users over a period of 30 days from an anonymous network operator. 
In our traces, each record contains information fields
such as user hashed ID, truncated-URL, download size, upload size and
timestamp. Note that our dataset is not restricted to any 
particular domain or limited to a small set of volunteer users. Being
a network-side dataset, it is fairly large and diverse in terms of the domains and
the users covered. The flip side of this is that it 
is also very noisy -- it contains not just the URLs that a user types in
his browser, but also all the redirects, secondary URLs
(pictures, embedded videos etc.) and
unintentional data (scripts, analytics, advertisement, spam etc.). 

\begin{figure*}[tb]
\begin{center}
\centering
\begin{adjustbox}{center}
  \centering
  \begin{subfigure}[t]{0.42\textwidth}
    \centering
    \includegraphics[scale=0.22]{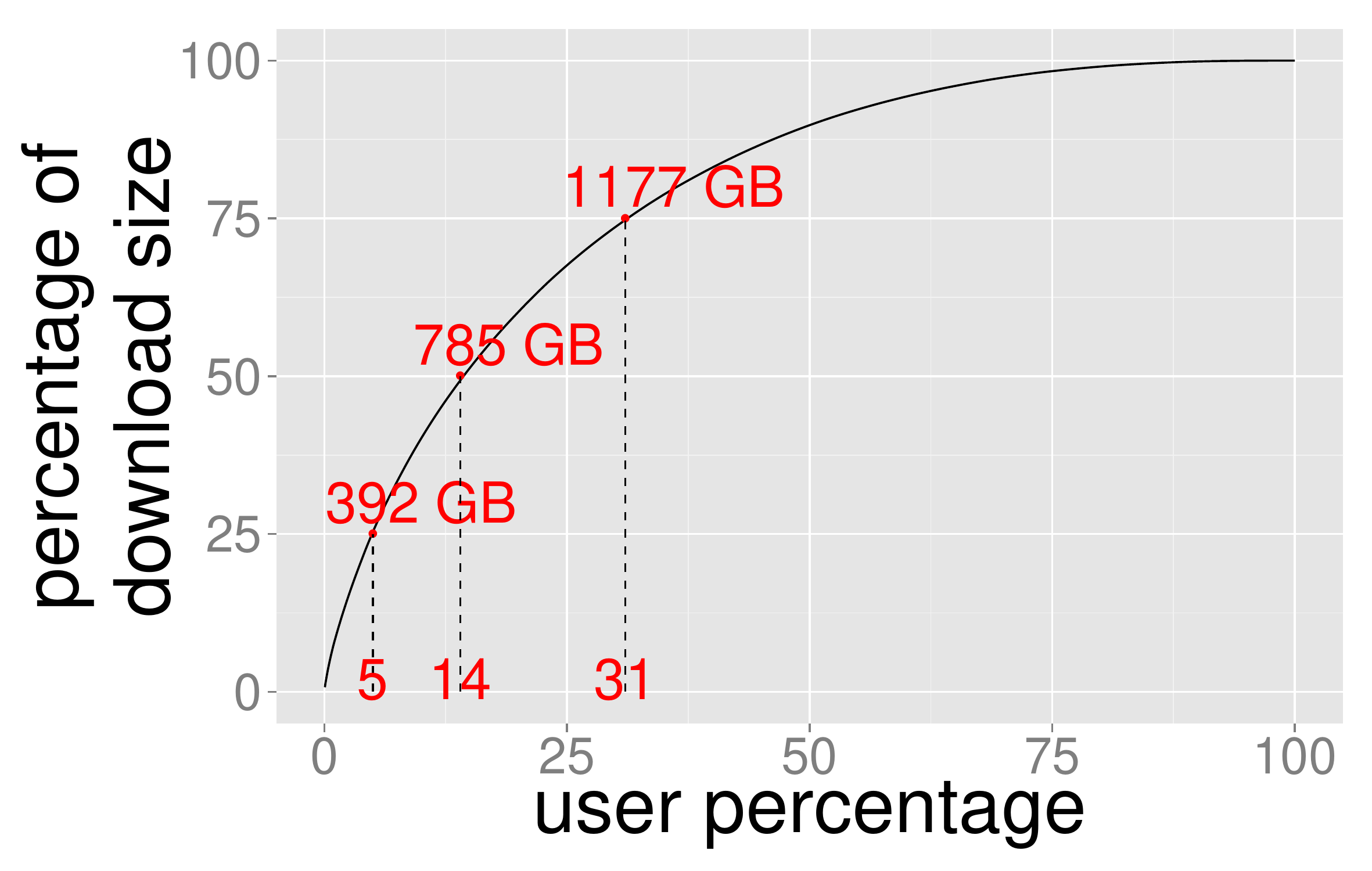}
    \subcaption{}
    \label{fig-do:sumstat_A}
  \end{subfigure}
  \begin{subfigure}[t]{0.42\textwidth}
    \centering
    \includegraphics[scale=0.22]{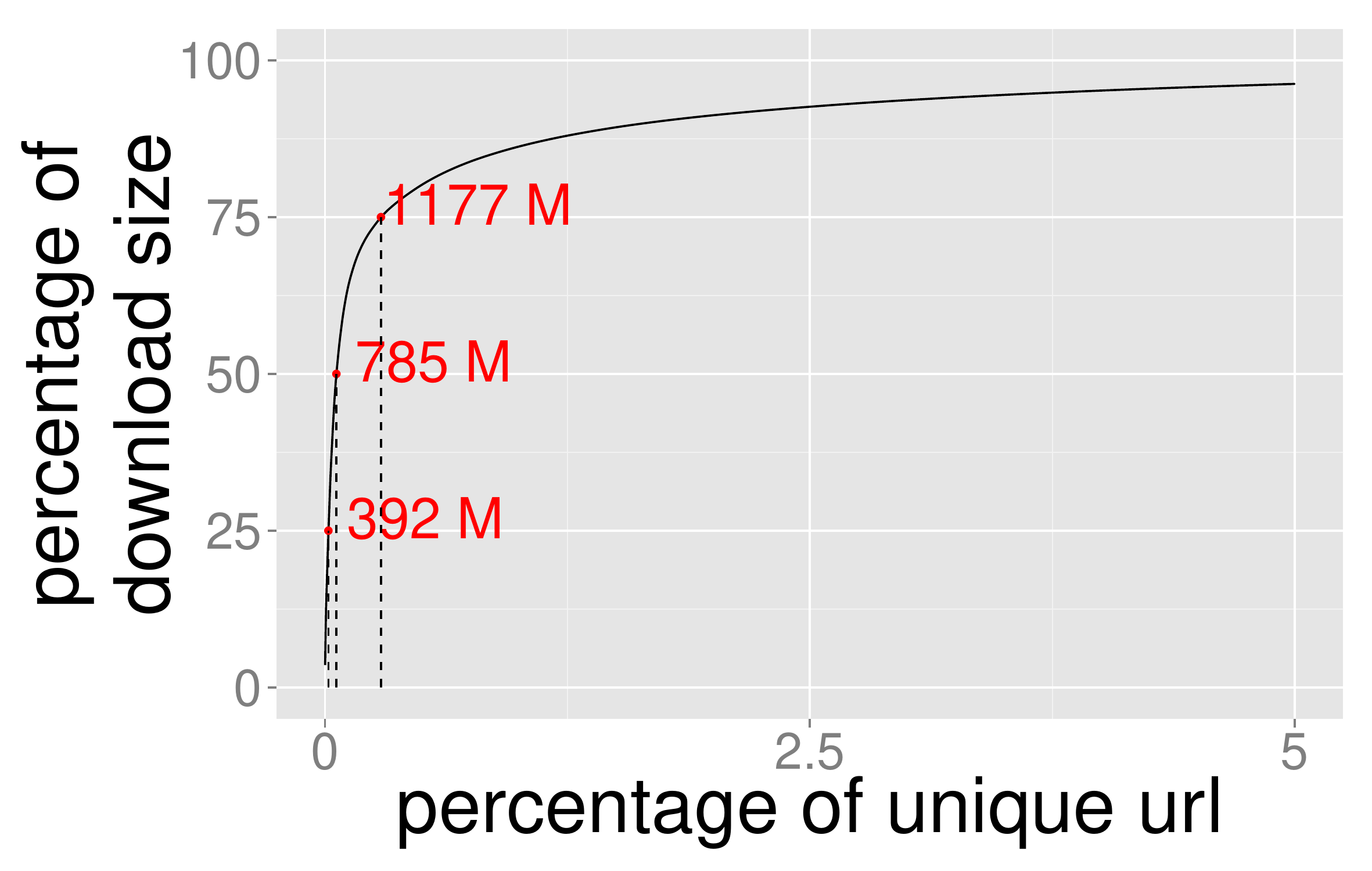}
    \subcaption{}
    \label{fig-do:sumstat_B}
  \end{subfigure}
  \begin{subfigure}[t]{0.42\textwidth}
    \centering
    \includegraphics[scale=0.22]{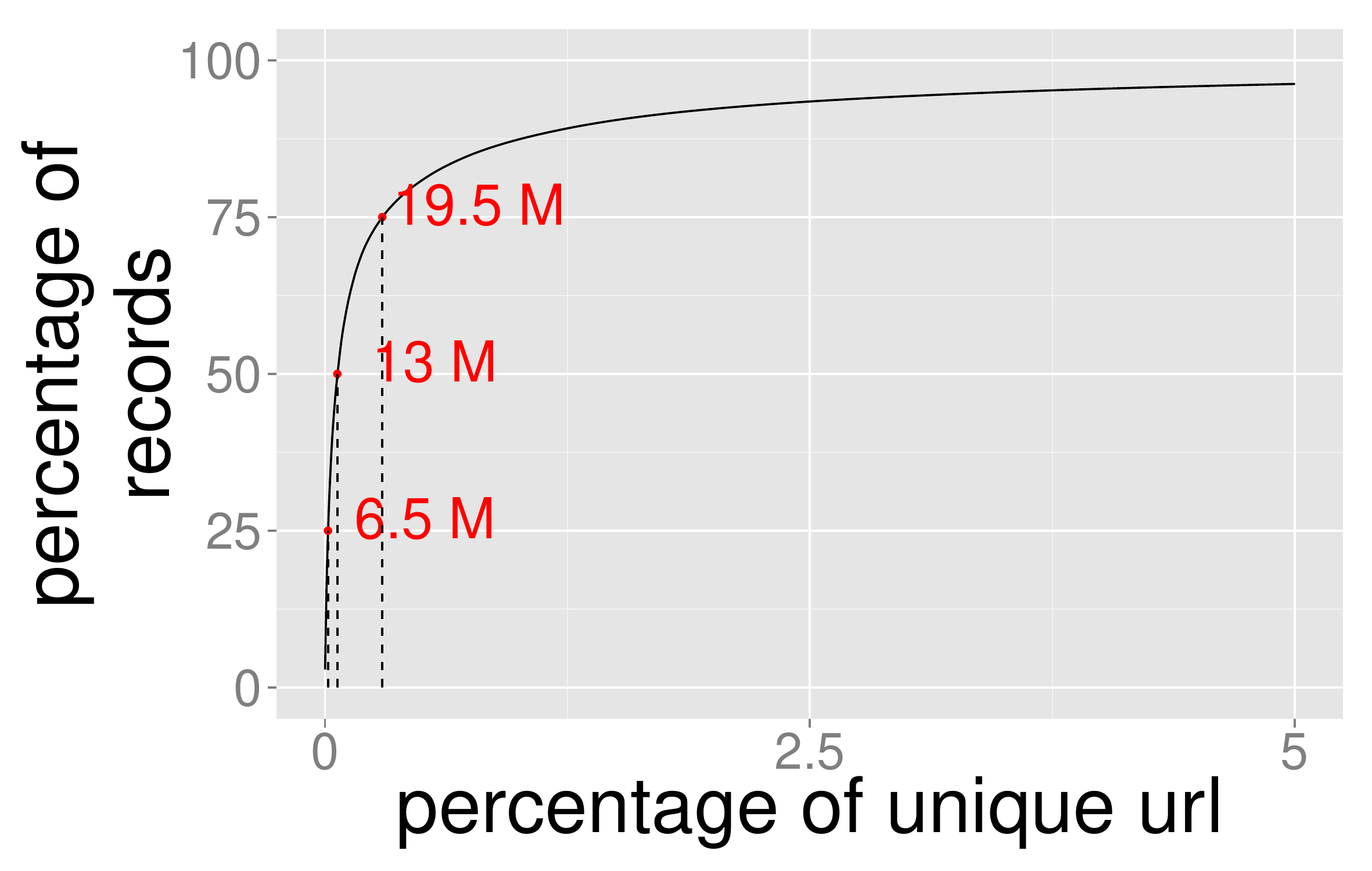}
    \subcaption{}
    \label{fig-do:sumstat_C}
  \end{subfigure}
\end{adjustbox}
\caption{Summary statistics for the traffic trace of a randomly chosen
  batch of $2000$ users.}
\label{fig-do:sumstat}
\end{center}
\end{figure*}

\para{Variation in Total Traffic.} We first observe that there is a
significant variation in the HTTP traffic generated by different users. 
For instance, the number of HTTP
records ranges from low tens for some users to tens of thousands for
other users, over the 30 day period of study. In fact, a majority of
HTTP download traffic ($75\%$) is generated by just $31\%$ of user
(Figure \ref{fig-do:sumstat_A}). We observe even more skewed distribution
for the traffic in terms of the generating activity domain. Less than
$0.5\%$ of domains generate $75\%$ of traffic in terms of download
size (Figure \ref{fig-do:sumstat_B}) and HTTP record counts
(Figure \ref{fig-do:sumstat_C}). Note that even though a large majority
($99.5\%$) of URLs together constitute only a small portion ($25\%$)
of the traffic, these less popular URLs are more likely to characterize the
unique features of different users and therefore, they play 
a critical role in differentiating 
user specific behavior. Thus, it is vitally important to correctly
classify these URLs into those that represent the user activities and
those that don't.

\begin{figure}[tb]
\centering
\begin{adjustbox}{center}
  \begin{subfigure}[t]{0.65\textwidth}
    \centering
    \includegraphics[scale=0.22]{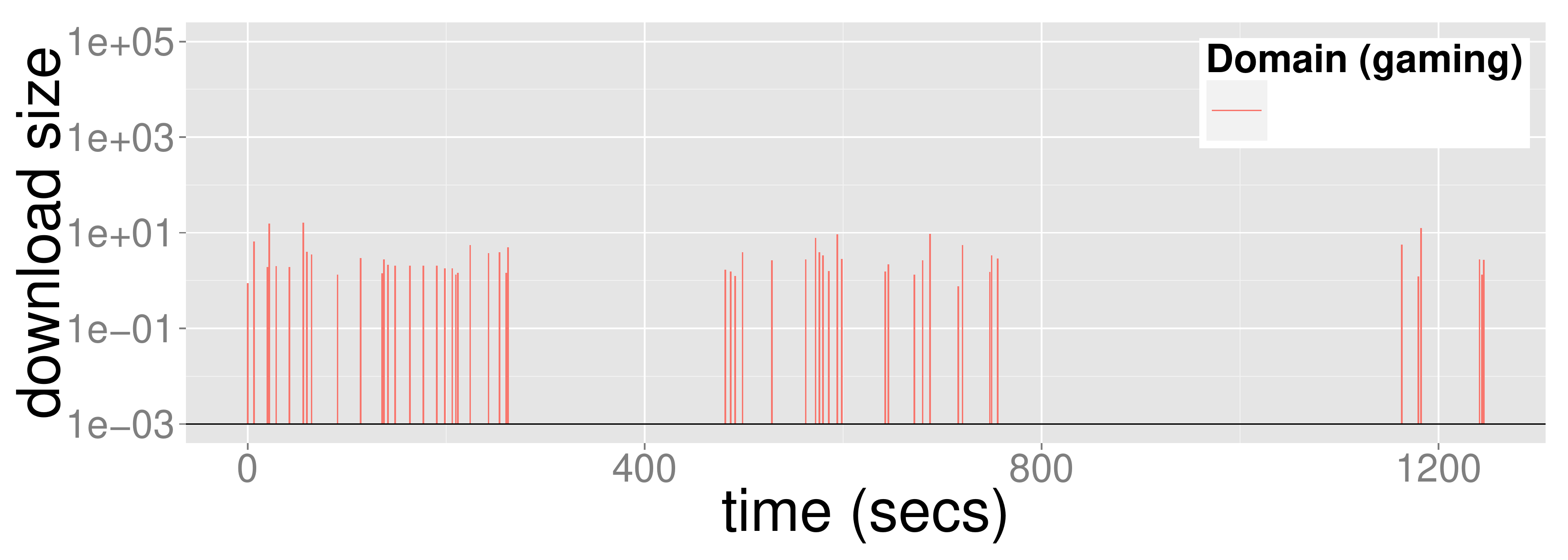}
    \subcaption{}
    \label{fig-do:timestamped_records_A}
  \end{subfigure}
  \begin{subfigure}[t]{0.65\textwidth}
    \centering
    \includegraphics[scale=0.22]{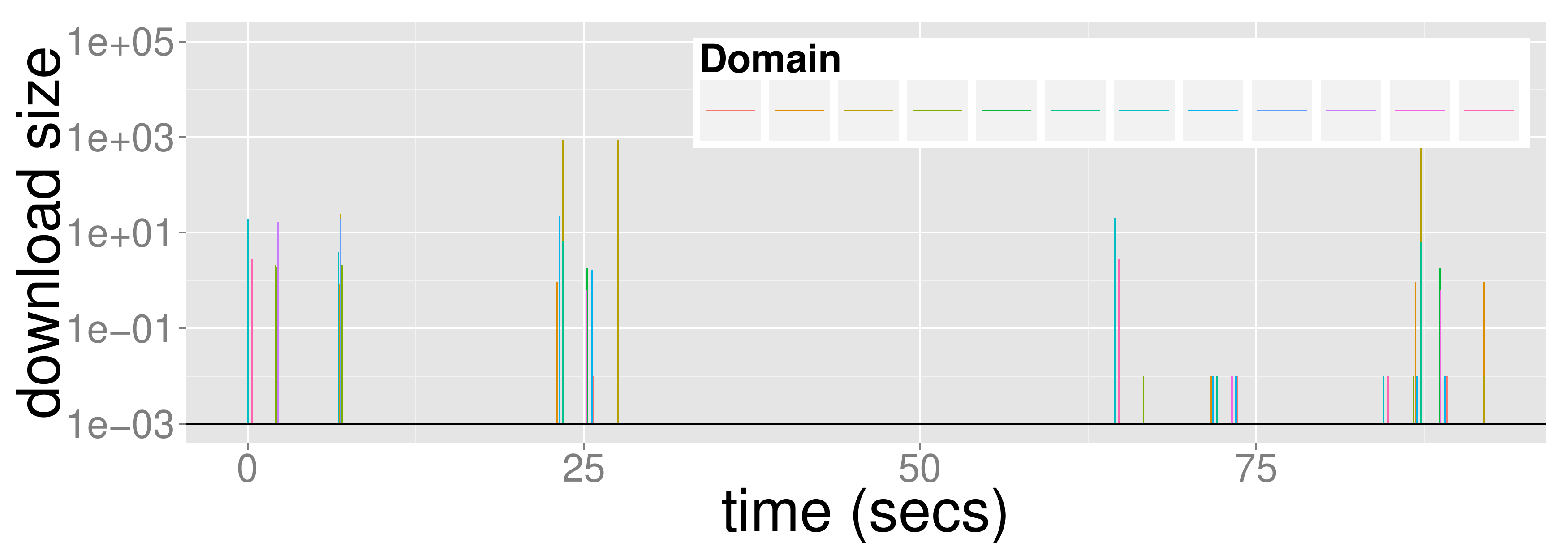}
    \subcaption{}
    \label{fig-do:timestamped_records_B}
  \end{subfigure}
\end{adjustbox}
\caption{Snapshot of timestamped HTTP records with download size for
  two users, showing a significant variation in amount of
  non-representative traffic.}
\label{fig-do:timestamped_records}
\end{figure}

\para{Variation in Type of Traffic.} Even among the users
with similar total traffic, the kind of web activities and the
fraction of non-representative URLs in the traffic trace
varies considerably between the users. For instance, 
Figure \ref{fig-do:timestamped_records} shows the web trace snapshot of two
users, illustrating two different activity patterns. Different colored
segments in this figure represent traffic from different 
domains, which can be either representative or
non-representative. The  trace of the first user 
(Figure \ref{fig-do:timestamped_records_A}) has only
one domain, i.e. gaming, and in fact, repeated records from a single
URL for more than 1300 seconds. For this user, there is no non-representative
traffic to filter out. However, the web browsing activity of
another user shown in Figure \ref{fig-do:timestamped_records_B} alternates
between a large number of domains (scripts, multimedia, HTML CSS,
advertisements, analytics etc.) in less than 100 seconds, even
though he/she is browsing a single web-page during this time. This variation in activity
patterns is reflected in download size, inter-arrival time as well as
number of HTTP records. In addition, the timestamp patterns of HTTP
records also varies significantly from one user to another (see
Figure \ref{fig-do:timestamped_records}). 

\begin{figure}[tb]
\centering
\begin{adjustbox}{center}
  \begin{subfigure}[t]{0.65\textwidth}
    \centering
    \includegraphics[scale=0.22]{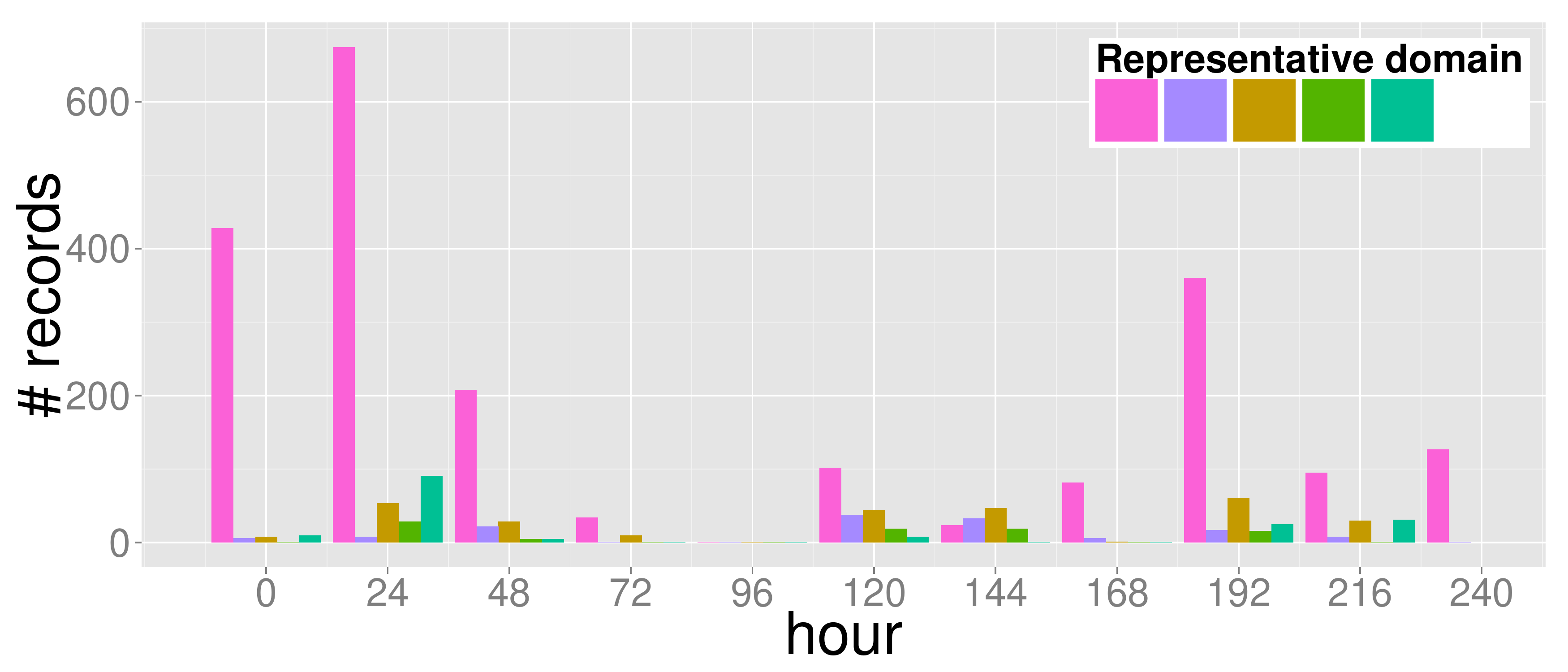}
    \subcaption{}
    \label{fig-do:daily_rec_count_A}
  \end{subfigure}
  \begin{subfigure}[t]{0.65\textwidth}
    \centering
    \includegraphics[scale=0.22]{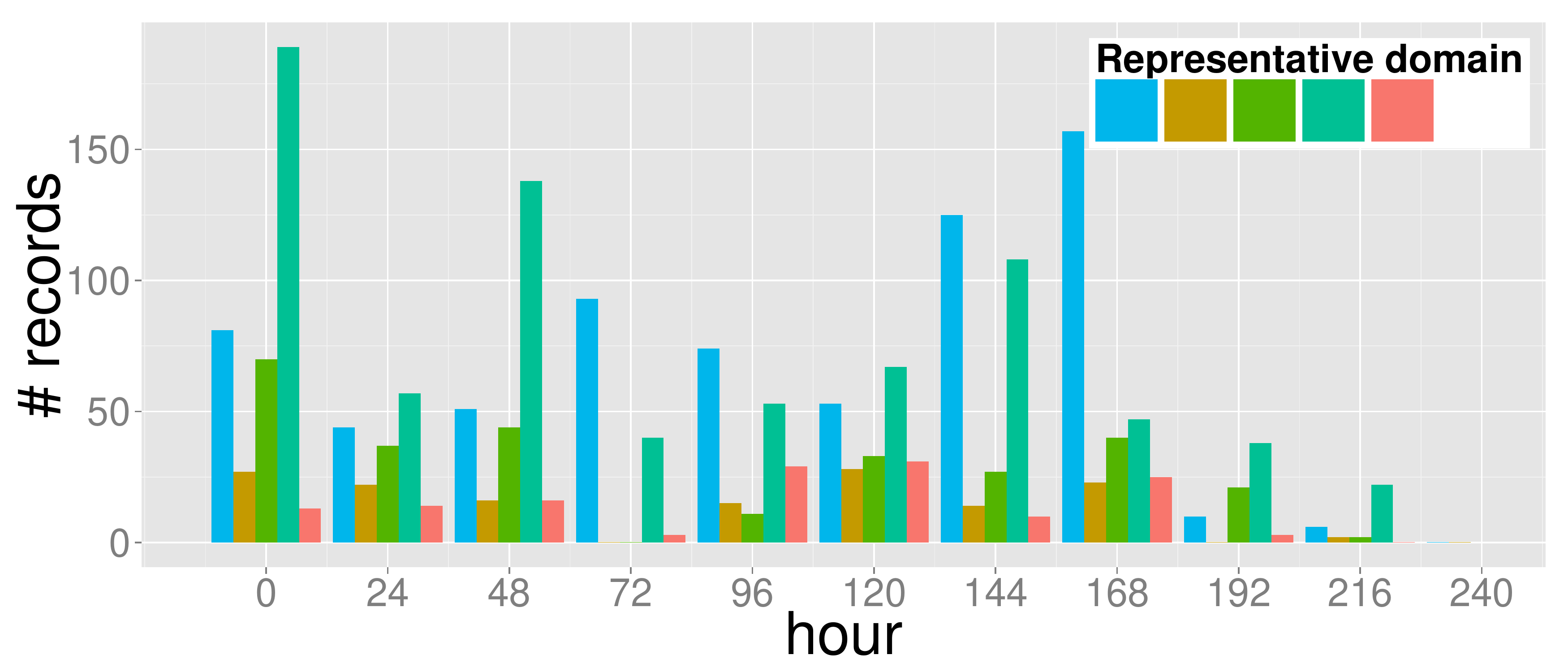}
    \subcaption{}
    \label{fig-do:daily_rec_count_B}
  \end{subfigure}
\end{adjustbox}
\caption{Daily record count of top $5$ user activities for two users; 
different colored bars represent different user activities such 
as reading, searching, gaming.}
\label{fig-do:daily_rec_count}
\end{figure}

\para{Variation in User Behaviors. } We also observe that 
there is a significant variation between
different users in terms of the activities
themselves. To summarize the aggregated variation
of the top-$k$ domains of both representative and non-representative
traffic, we use the following global entropy-based metric to measure this variation: 

\begin{align}
\label{eqn-do:aggvar_topkdom}
S_{a} &= \frac{-\sum_i \frac{n_i}{nk}\log(\frac{n_i}{nk}) - \log(k)}{\log(nk)-\log(k)} ,
\end{align}
where $n_i$ is the number of times that URL $i$ appears in the top $k$
domains, satisfying $\sum_i n_i = nk$. By
Equation 1
, the variation metric $S_a$ is maximized
at $1$ when all users have different non-overlapping top $k$ domain
set and is minimized at $0$ when all users have the same non-ordered
top $k$ domain set. For the web trace data, $S_a$ is $0.484$ with
$1455$ distinct domains from among the top $k=5$ domains for the 2000
users. The discovered $S_a$ value suggests that there 
is a significant variation in the top activities among the different users. 
We show this intuition graphically in 
Figures \ref{fig-do:daily_rec_count_A} and \ref{fig-do:daily_rec_count_B}, 
where we depict the activity
variation of two users over time. For this figure, we filtered 
out the non-representative
domains manually, selected top $5$ representative domains for each
user according to the number of HTTP
records. Figure \ref{fig-do:daily_rec_count} presents the daily record
counts for each representative domain and demonstrates both the
temporal and activity variations in terms of activity types and the 
magnitude across two randomly selected users.


\para{Summary.} These above variational statistics imply that the methods to 
extract features for separating noise from the representative
URLs have to adapt to changing user patterns. In particular, the
variation in the total traffic and the timestamp patterns necessitates
user-adaptive solutions that we explore in the next sections. 

\section{Our Methodology}
\label{sec:methodology}
In this section, we present an overview of our methodology to automatically classify the web-domains into 
those that represent the user activities and those that don't.
The key feature of this methodology is the usage of novel features derived from the burst decomposition of a user's web-trace that improves the accuracy of the classification, offsetting the loss due to URL truncation.

The main intuition behind our methodology is that a user's browsing activity consists of several data bursts. These data bursts correspond to micro user actions, such as a web click or a chat reply. In each burst, there are some URLs representing the user activity intermixed with other unintentional web-traffic such as advertisements, web-analytics 
 etc and secondary URLs corresponding to multimedia associated with the representative URL. Our statistical methodology decomposes the web-trace back into its constituent data bursts. It then leverages specialized features from data bursts (e.g., the position of a URL in a data burst, the number of unique URLs in a data burst, burst duration, burst download size etc.) to segregate the representative web-domains from the remaining web-domains. In Section \ref{sec:domainclassification}, we show that the usage of features derived from data burst help in significantly improving the accuracy of the segregation task.

A key challenge in our methodology is the decomposition of the web-trace into data bursts. As highlighted already in Section \ref{sec:variation}, there is a considerable variation in the traffic patterns of different users. We found that even the distribution of inter-arrival time of HTTP records is very different for different users.  
This makes it particularly difficult to model these data bursts and to find good thresholds to decompose the web-trace into data bursts. We solve this problem by having different thresholds for different users and ensuring that the threshold computing function is robust with respect to the distribution shape. 
This is achieved using a novel technique to generate thresholds for each user that adapts to any distribution of inter-arrival time.

\section{Inter-arrival Time Distribution Models}
\label{sec:inter-arrival}

In this section, we study the inter-arrival time of HTTP records with
a view to finding good thresholds that will decompose a user's
traffic-trace into burst of records that represent micro user actions.



As described in Section \ref{sec:methodology}, the key 
concept behind burst is that when a user
performs a micro action like web click, chat reply etc., it not only
generates many HTTP records related to the representative activity, but also a
large number of secondary records such as advertisements, web
analytics, webscripts etc. These records are all intermixed. When the
user completes the current micro-action, e.g. reading the current web
page, and starts a new one, e.g. opening the next page, a new burst is
generated with its associated records. So, the observed inter-arrival
time records are the combined results of within-burst and out-of-burst
records. However, we expect that the within-burst HTTP records are
closer together and the out-of-burst records are far apart in
time. By computing an appropriate \textit{separation threshold} on the
inter-arrival time, we aim to decompose the traffic into its
constituent bursts.


Since traffic patterns and the inter-arrival time distributions for
different users are very different, we
can't expect a global threshold to work well for all users. Instead,
we compute a different threshold for each user specific to his/her traffic
patterns. If the difference between the time-stamp of a record and its
predecessor is greater than the computed 
\textit{separation threshold} for that user, the
record marks the beginning of a new burst. Otherwise, the record
belongs to the burst of its predecessor.

To learn the \textit{separation threshold} for each user, our first approach
is to learn the probability density function of inter-arrival time for
the users. By computing the best-fitting parameters for
this density function for each user and defining the separation threshold as a function
of those parameters, we can decompose the traffic trace for each user into its
constituent bursts.

We modeled personalized
inter-arrival time distributions by exploring different density
functions, such as exponential distribution, pareto distribution and
mixtures and concatenations of these distributions (details provided
in Appendix \ref{sec:model-inter-arrival}). From the analysis, we
found that even these general density functions are not
flexible enough to accommodate highly varied and personalized
inter-arrival time of different users. Thus, we concluded that even
though this formalism is principled, there is a need for a more  
robust technique to separate within-burst and out-of-burst records, 
that is \emph{independent} of the personalized distribution shape of a user.

\section{Burst Decomposition Using Adaptive Thresholds}
\label{sec:burstdetection}

In this section, we propose a robust burst decomposition algorithm
that is independent of the distribution shape. Our technique only relies on the general
characteristics of the inter-arrival time distribution observed in
Appendix \ref{sec:model-inter-arrival}, but not on any specific
model. The only characteristic of the inter-arrival time 
distribution that we use is that there is a within-burst
component with high arrival-rate of records (and small inter-arrival
time), an 
out-of-burst component with low arrival-rate (high inter-arrival time) forming a long tail and that these two
components are separable with a threshold. Our aim in this section is
to have a threshold that adapts itself to any inter-arrival time
distribution, subject to this general property.

We first observe that an optimal threshold $\tau^\star$ is expected to lay in
a low probability range and should satisfy the following conditions:
\begin{itemize}
\item
$\forall$ $x < \tau^\star$, $p(x)$ should, generally, be high and show the presence of bursts
\item
$\forall$ $x >= \tau^\star$, $p(x)$ should, typically, have low values and imply user inactivity periods 
\end{itemize}

In order to satisfy the above conditions,  $\tau^\star$ has 
to intercept the minimum $x$ point where the probability density 
function of inter-arrival time distribution decays
to fairly close to zero and the density of $p(x)$ values beyond $\tau^\star$ is minimal. 

However, to quantitatively measure the significance of each $p(x)$ value, 
we need a \textit{scalar indicator} that would determine when a $p(x)$ 
value is minimal. This approach would suffer from the selection of 
a global scalar indicator that would fail in detecting the 
intrinsic variations of the density proportion between the within 
bursts and out-of bursts components for different users. 

Therefore, instead of using this approach of quantifying $p(x)$, 
we leverage the conditional density, i.e. $\frac{p(x)dx}{\int_{0}^{x}p(y)dy}$, 
to determine $\tau^\star$.
Note that, $\frac{p(x)dx}{\int_{0}^{x}p(y)dy}\approx \frac{c_i}{\sum_{j=1}^i c_j}$ 
which is the probability that a time sample belongs to bin $i$, conditioning on the
fact that it belongs to a bin less than or equal to $i$: $k_i = \Pr(x
\in b_i|x \in b_{1:i})$. In other words, $k_i$
measures the contribution of the current bin to the accumulated
probability.

  \begin{algorithm}[t]
  \caption{Burst Decomposition}
  \label{alg-bd:burstdetection}
    \begin{enumerate}
    \item Divide inter-arrival times into bins $b_i$ of length $l$ with counts $c_i$
    \item Calculate $k_i = \frac{c_i}{\sum_{j=1}^{i}c_j}$
    \item Find the smallest index $i^{\star}$ that $k_{i^{\star}+j}<p$ $\forall j=1..J$
    \item Define the inter-arrival threshold $\tau^{\star}=l \times i^{\star}$
    \item Group consecutive URLs with inter-arrival times $\tau < \tau^{\star}$ into bursts.
    \end{enumerate}
  \end{algorithm}

Our Algorithm \ref{alg-bd:burstdetection} searches for $\tau^\star$ 
by starting from the smallest value of the
inter-arrival time density such that the extended probability by
increasing decaying point is insignificant, compared to the
accumulated probability at that point (as captured by
$k_i$). Specifically, the threshold $\tau^{\star}$ is found 
when the contributions of $J$ consecutive bins are less than a 
predefined probability, for a pre-specified parameter $J$.  

The burst decomposition algorithm will group all the records with 
inter-arrival time less than the obtained $\tau^\star$ into actual bursts.

In the next section, we provide evidence that this algorithm detects
meaningful bursts that significantly improve the classification
accuracy in identifying the domains that represent user activities.

\begin{figure}[!ht]
\centering
\includegraphics[scale=0.22]{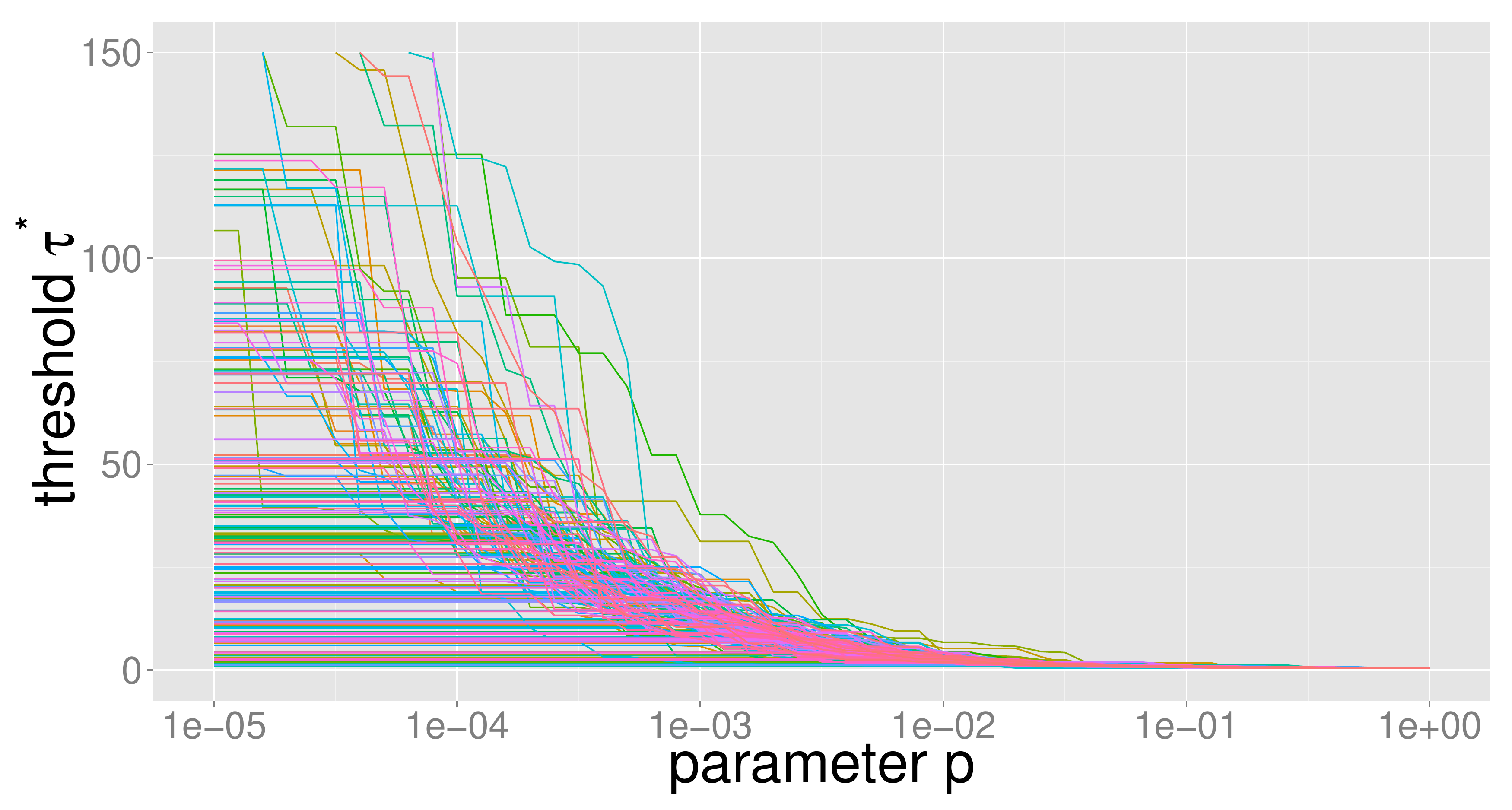}
\caption{Calibrating parameter $p$ by examining the variation of user behaviors through the threshold $\tau^\star$.}
\label{fig-bd:threshold_peruser}
\end{figure}

We estimate the values of the scalar indicator, $p$, used in 
the Algorithm \ref{alg-bd:burstdetection} based on an analysis of the 
corresponding $\tau^\star$ values across all users. In 
Figure \ref{fig-bd:threshold_peruser} we only report the $\tau^\star$ 
behaviors of $200$ users as representative of entire $\tau^\star$ 
values computed across all users. It is easy to notice that for 
$p=0.01$ the $\tau^\star$ would range from $1$ to $10$ seconds, 
which is a reasonable range to separate inter-arrival time values 
between within burst and out-of bursts for activities such as web browsing, reading,
shopping, etc. Hence, this value of $p$ was used in our experimental analysis.

\begin{figure*}[t]
\centering
\begin{adjustbox}{center}
  \begin{subfigure}[t]{0.42\textwidth}
    \centering
    \includegraphics[scale=0.22]{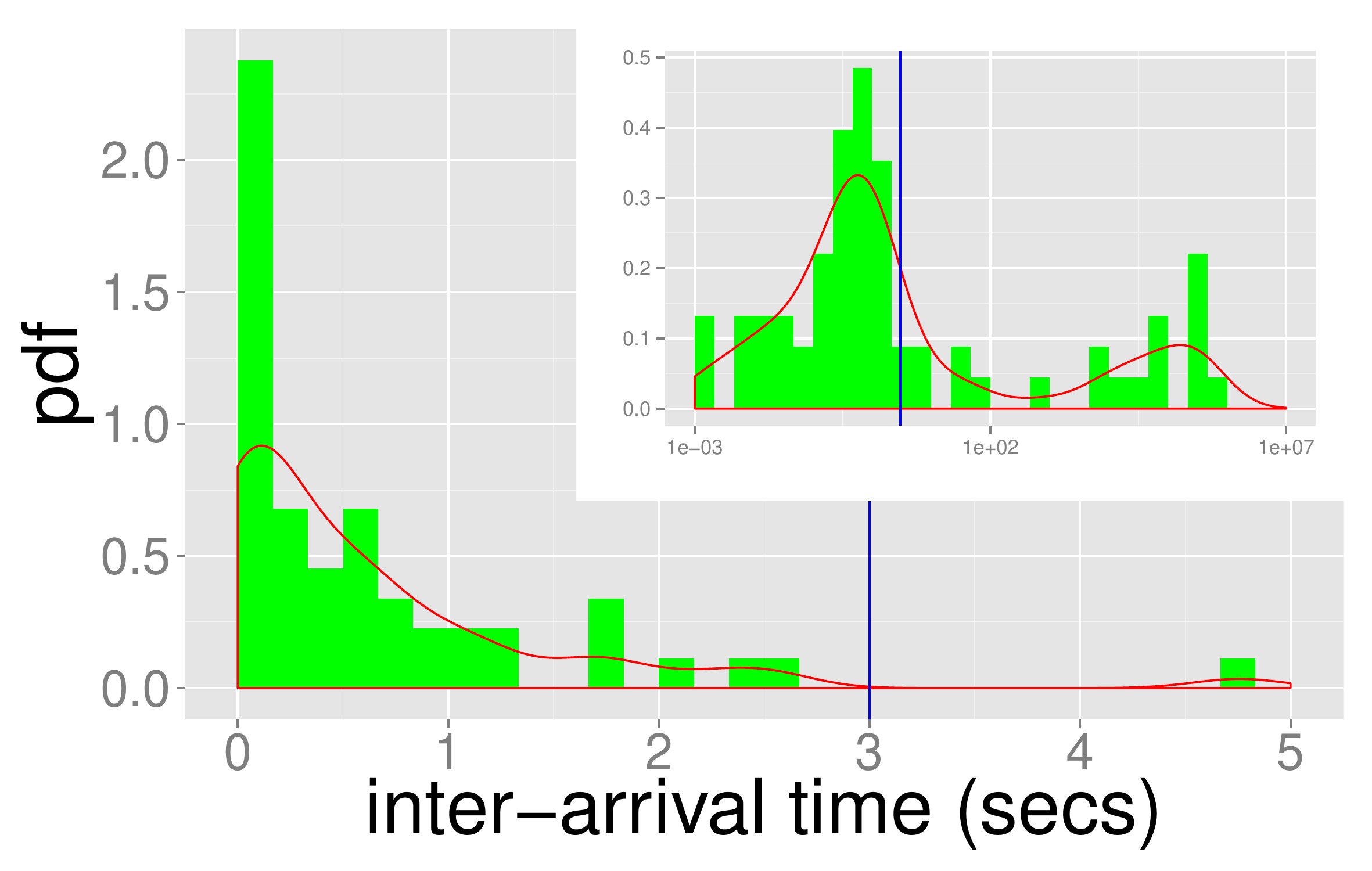}
    \subcaption{Low activity user: $71$ records.}
    \label{fig-bd:bd_peruser_A}
  \end{subfigure}
  \begin{subfigure}[t]{0.42\textwidth}
    \centering
    \includegraphics[scale=0.22]{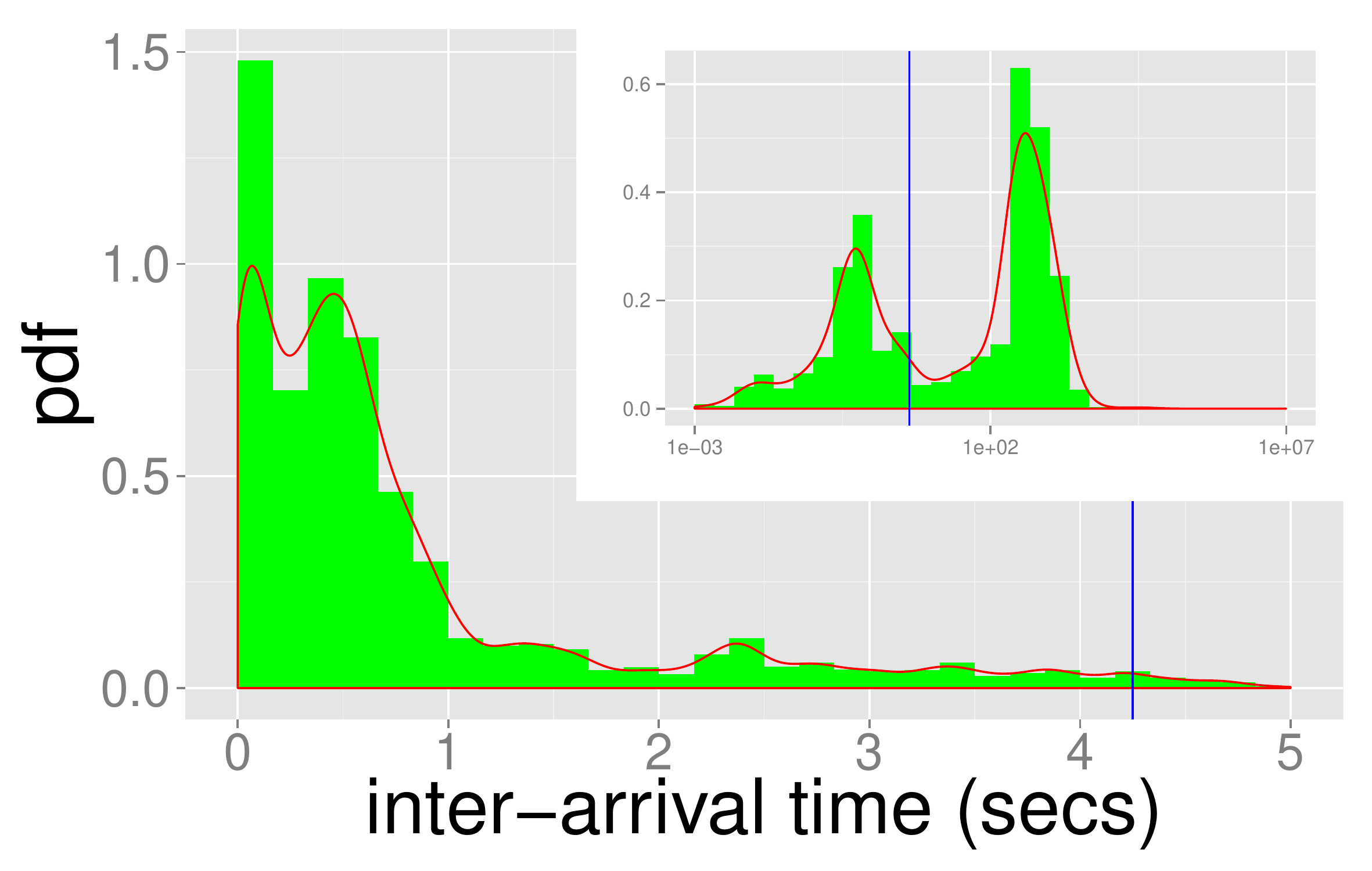}
    \subcaption{Medium activity user: $6764$ records.}
    \label{fig-bd:bd_peruser_C}
  \end{subfigure}
  \begin{subfigure}[t]{0.42\textwidth}
    \centering
    \includegraphics[scale=0.22]{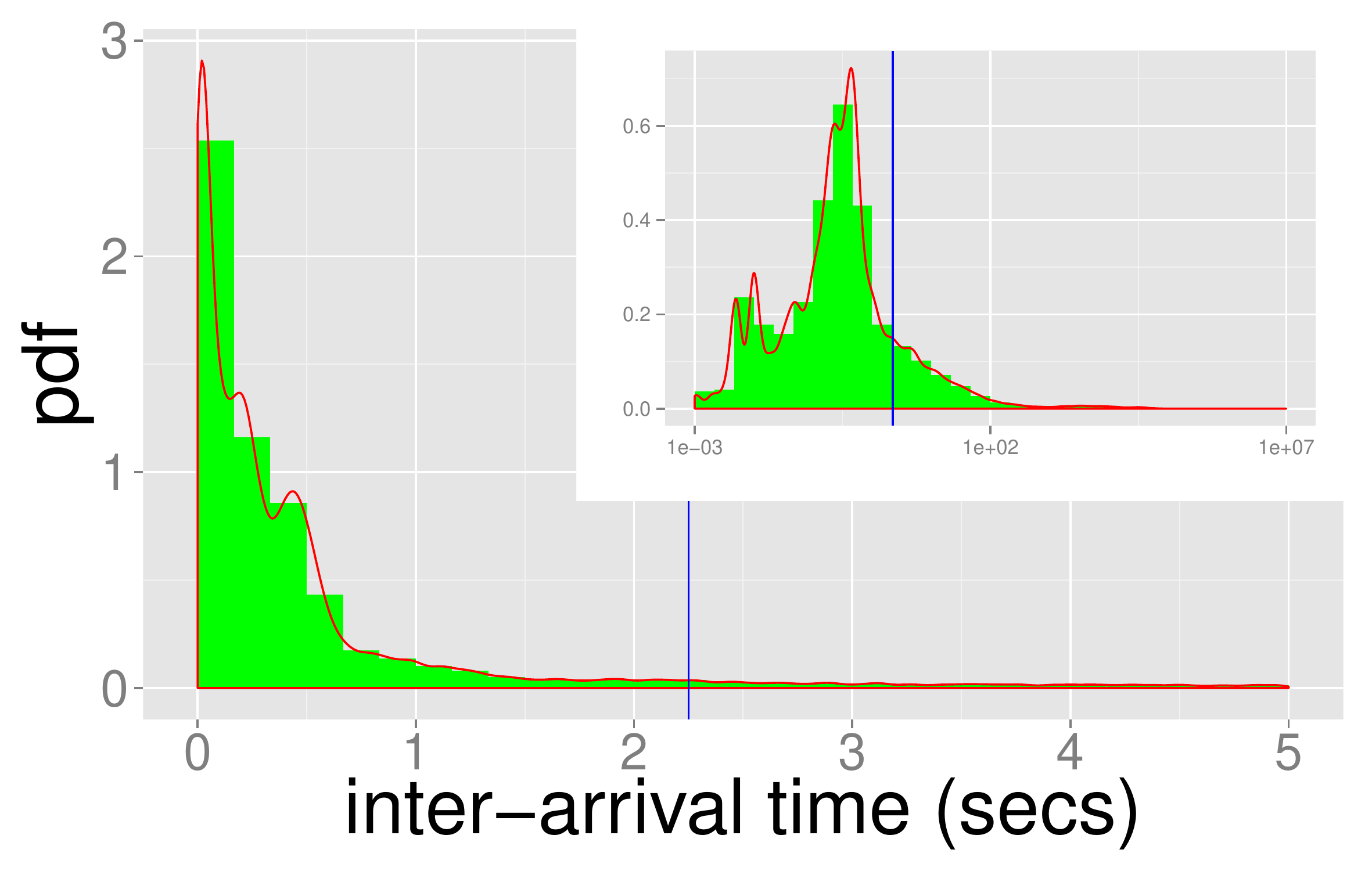}
    \subcaption{High activity user: $35116$ records.}
    \label{fig-bd:bd_peruser_B}
  \end{subfigure}
\end{adjustbox}
\caption{User dependent $\tau^\star$ obtained from burst decomposition algorithm.}
\label{fig-bd:bd_peruser}
\end{figure*}

Next, we examine the results of our algorithm with respect to 
users with substantially different behaviors. In particular, we 
leverage the three users examined in Figure \ref{fig-bd:iatden}. Even though the
distribution shapes and the number of records characterizing these
three users are very different, the algorithm successfully finds 
a user specific $\tau^\star$ as shown in Figure \ref{fig-bd:bd_peruser}.

\section{Domain classification}\label{sec:domainclassification}

In this section, we describe our classification model for identifying
the representative URLs and show that it is possible to achieve very
high accuracy for this task even with truncated URLs. Features extracted
from the burst decomposition presented in Section \ref{sec:burstdetection} play a
crucial role in significantly improving the accuracy of  our
classification model.

\para{Classification Model Formalization.} 
We use a logistic regression model for the domain classification problem. 
Our model for logistic regression is as follows:

\begin{align}
y_i &\sim \Bernoulli(q_i),\\
\ln(\frac{q_i}{1-q_i}) &= \mu_i = \beta_0 + \sum_l \beta_l x_{i,l},
\end{align}
where $y_i$ is the binary label ($y_i=1$ if URL $i$ is representative and 
$y_i=0$ otherwise) and $x_{i,l}$ is the specific classification feature 
that we derive from record-level and burst-level analysis in 
Sections \ref{sec:record_level} and \ref{sec:burst_level}. The 
representative probability $q_i$ is computed by the \textit{logistic function} 
on a linear predictor $\mu_i$ and all the parameters are estimated by the 
Iteratively Re-Weighted Least Squares (IRWLS) method \cite{Wood2006}.

The domain classification follows three steps. First, we manually 
label $400$ URLs into two classes: representative and non-representative 
domains. Second, we extract five sets of web traces generated out 
of $2K$ random users each, perform the burst decomposition and obtain 
aggregated measurements independently for each set. Finally, half of 
the labelled URLs of the first set are used in training the classifier, 
which is validated by the other half of the first set and the remaining 
four. We use five different sets to validate the robustness of our approach.

We demonstrate the accuracy of our classification approach in two steps. 
We first study the accuracy obtained by only
using the record-level features and ignoring the burst-level
features. Then, we show the improvements we gain by adding the burst-level 
features which are derived upon the detected bursts from our burst decomposition algorithm. 

\para{Record-level Features.} The key part of our modeling is feature
engineering, or identifying the right set of features to achieve a
high accuracy. For the record-level features, shown in 
Table \ref{tab-record:aggmeasurement}, we use the aggregated
measurements across all users and compute the quantile values $k$ by 
ranging $k$ from $5$ to $95$ with an increment step equal to $5$. These
features were carefully selected to achieve a high accuracy 
with record-level features. Specifically, for each record we collect the 
leading and following inter-arrival time and the upload and download size. 
These features are examined as covariates in our domain classification model.

\subsection{Accuracy with Record-level Features}\label{sec:record_level} 

\para{Accuracy.} As shown in Table \ref{tab-record:accuracy_BM}, 
the resultant accuracy with
the record-level features is quite poor. For the five sets of web
traces, the accuracy varies between 69.7\% and 72.9\%, implying that
around 30\% of the URLs are misclassified. 
Among the analyzed features we have discovered two particularly important:
$x_{\cdot,1}=(s_{r,d,75\%}-s_{r,d,25\%})$ and
$x_{\cdot,2}=s_{r,u,50\%}$ by the stepwise model selection procedure. 
The first is the difference between the $75$ and $25$ quantile 
statistics of the download size per domain and the second is 
the $50$ quantile statistic of the upload size. The estimated 
coefficients for this model is shown in Table \ref{tab-dc:logreg_sum_record}, 
implying that domains with small variation of download size and high 
value of upload size have higher chance of being representative domains. 
However they are the most relevant features at record-level, their 
discriminatory capacity still remains limited.

\begin{table}[tb]
\caption{Aggregated measurements at record-level where $\mathcal{R}(i)$ 
denotes the set of records containing domain URL $i$.}
\centering
\begin{tabular}{|c|l|}
\hline
\multicolumn{2}{|c|}{Record-level features (wrt $\mathcal{R}(i)$)}\\
\hline
\hline
$t_{r,l,k}$ & Quantile $k$ of the leading inter-arrival time\\
\hline
$t_{r,n,k}$ & Quantile $k$ of the next inter-arrival time\\
\hline
$s_{r,u,k}$ & Quantile $k$ of the upload size\\
\hline
$s_{r,d,k}$ & Quantile $k$ of the download size\\
\hline
\end{tabular}
\label{tab-record:aggmeasurement}
\end{table}

\begin{table}[tb]
  \centering
  \begin{adjustbox}{center}
    \begin{minipage}{.65\linewidth}
    \captionsetup{width=0.9\textwidth}
    \caption{AIC and classification accuracy with record-level features.}
    \label{tab-record:accuracy_BM}
    \centering
      \begin{tabular}{|c|r|r|r|r|r|r|r|}
      \hline
      & \multicolumn{1}{c|}{AIC} & \multicolumn{1}{c|}{BIC} & \multicolumn{5}{c|}{Classification accuracy on }\\
      & \multicolumn{1}{c|}{} & \multicolumn{1}{c|}{} & \multicolumn{5}{c|}{ 5 sets of 2K users each.}\\
      \hline
      $\mathcal{C_R}$ & $242.82$ & $252.72$ & $69.7$ & $72.9$ & $71.0$ & $70.8$ & $71.4$ \\
      \hline
      \end{tabular}
    \end{minipage}%
    \begin{minipage}{.65\linewidth}
    \captionsetup{width=0.9\textwidth}
    \caption{Estimated values ($\beta_l$), standard deviation 
    ($\sigma_{\beta}$), $p$-values and significance (SIG) for 
    logistic regression model with record-level features.}
    \label{tab-dc:logreg_sum_record}
    \centering
      \begin{tabular}{|c|r|r|r|r|}
      \hline
      Feature & \multicolumn{1}{c|}{$\beta_l$} & \multicolumn{1}{c|}{$\sigma_{\beta}$} & \multicolumn{1}{c|}{$p$-value} & \multicolumn{1}{c|}{SIG}\\
      \hline
      Intercept  & $-1.41$ & $0.38$ & $1.8e-4$ & ***\\
      \hline
      $s_{r,d,75\%}-s_{r,d,25\%}$  & $-0.03$ & $0.01$ & $4.7e-4$ & ***\\
      \hline
      $s_{r,u,50\%}$  & $0.86$ & $0.26$ & $1.0e-3$ & ***\\
      \hline
      \end{tabular}
    \end{minipage} 
  \end{adjustbox}
\end{table}

%

\subsection{Accuracy with Burst-level Features}\label{sec:burst_level}
In this section, we show how the accuracy improves with 
features measured at burst-level.

\begin{figure*}[tb]
\centering
\begin{adjustbox}{center}
  \begin{subfigure}[t]{0.65\textwidth}
    \centering
    \includegraphics[scale=0.22]{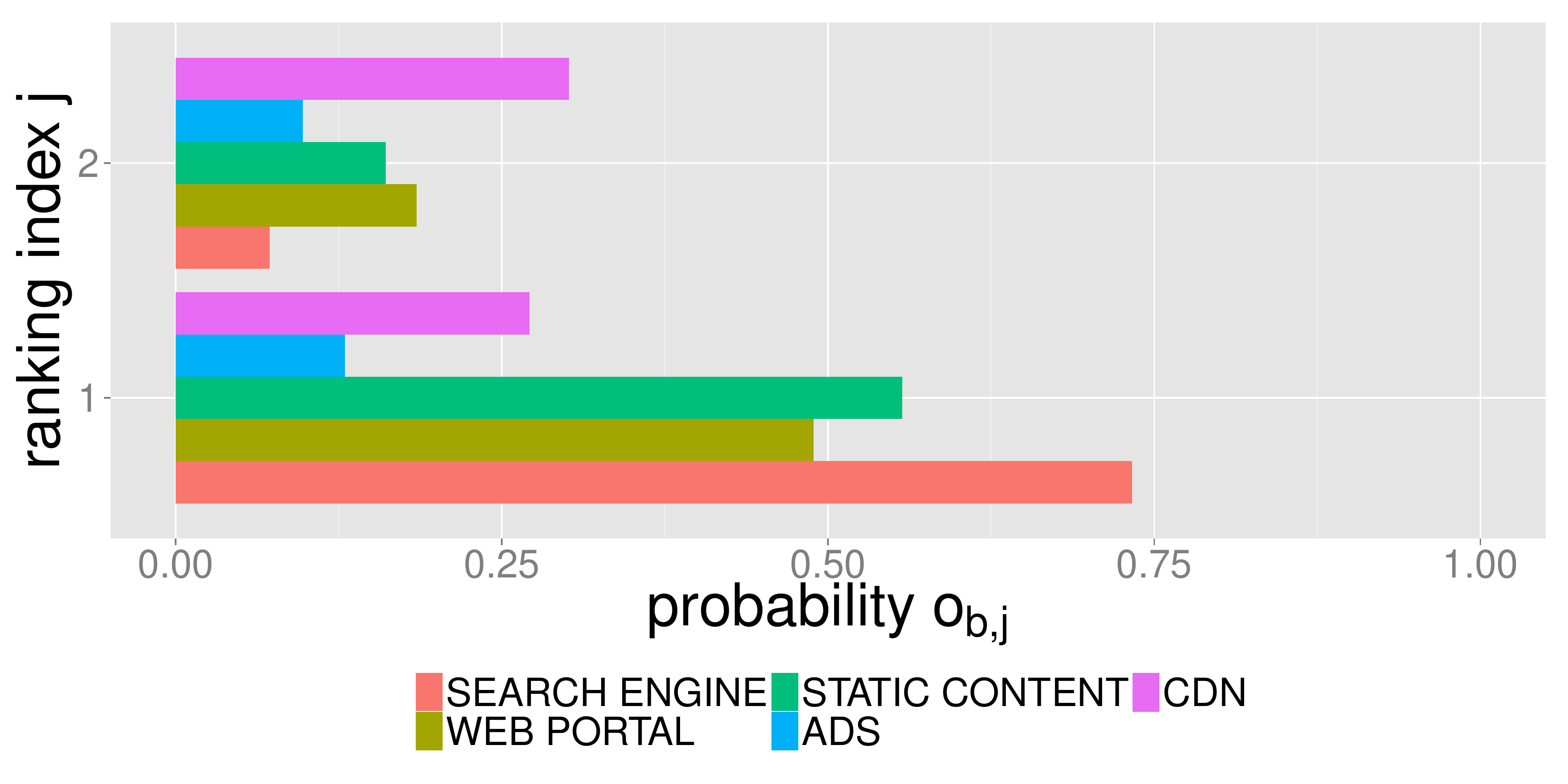}
    \subcaption{}
    \label{fig-bd:measurement_order}
  \end{subfigure}
  \begin{subfigure}[t]{0.65\textwidth}
    \centering
    \includegraphics[scale=0.22]{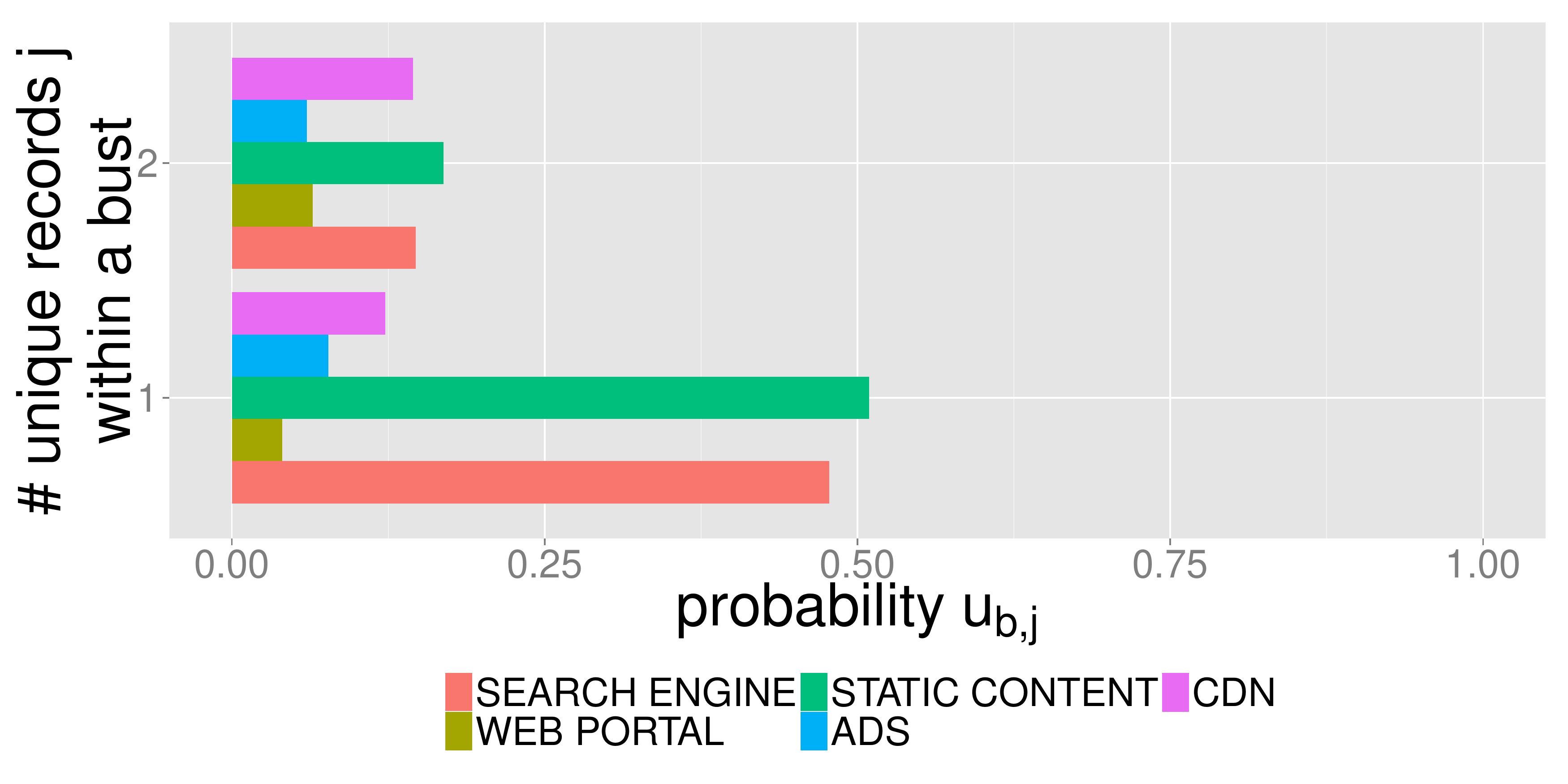}
    \subcaption{}
    \label{fig-bd:measurement_unirecnum}
  \end{subfigure}
\end{adjustbox}
\caption{The probability $o_{b,j}$ that a URL is ranked $j$ in its burst
  and the probability $u_{b,j}$ that there are  $j$ records in the
  burst containing the URL ($j=1:2$) for different domains.}
\label{fig-bd:measurement}
\end{figure*}

\para{Burst-level Features.} By leveraging the burst decomposition 
algorithm, we segment our web traces in a series of consecutive bursts and we
measure burst-specific characteristics. Specifically, for 
each URL $i$, we choose a list of aggregated
measurements, shown in Table \ref{tab-bd:aggmeasurement}, where
$\mathcal{B}(i)$ denotes the set of bursts containing URL $i$. 
We observe that two
burst features, i.e. $o_{b,j}$ and $u_{b,j}$ ($j=1:2$), in
Table \ref{fig-bd:measurement} are particularly important in
improving the domain classification results. The first measure, 
i.e.  $o_{b,j}$, describes the probability that a URL is ranked $j$ 
in its burst and the second, i.e. $u_{b,j}$, quantifies 
the probability that there are  $j$ unique domains in the burst 
containing the URL ($j=1:2$). Similar to record-level
features, these aggregated
measurements are examined as covariates in our domain classification
model.

\begin{table}[tb]
  \centering
  \begin{adjustbox}{center}
    \begin{minipage}{.65\linewidth}
    \captionsetup{width=1\textwidth}
    \caption{Aggregated measurements at burst-level where $\mathcal{B}(i)$ 
    denotes the set of bursts containing URL $i$.}
    \label{tab-bd:aggmeasurement}
    \centering
      \begin{tabular}{|c|l|}
      \hline
      \multicolumn{2}{|c|}{Burst-level features (wrt $\mathcal{B}(i)$)}\\
      \hline
      \hline
      $o_{b,j}$ & The probability that URL $i$ is ranked $j$-th\\
      & ($j=1:9$)\\
      \hline
      $u_{b,j}$ & The probability that a burst containing\\
      & URL $i$ has $j$ unique URLs ($j=1:9$) in the burst\\
      \hline
      $d_{b,k}$ & Quantile $k$ of burst duration\\
      \hline
      $t_{b,l,k}$ & Quantile $k$ of the leading inter-arrival time\\
      & of a burst\\
      \hline
      $t_{b,n,k}$ & Quantile $k$ of the next inter-arrival time of\\
      & a burst\\
      \hline
      $s_{b,u,k}$ & Quantile $k$ of the burst upload size\\
      \hline
      $s_{b,d,k}$ & Quantile $k$ of the burst download size\\
      \hline
      \end{tabular}
    \end{minipage}%
    \begin{minipage}{.65\linewidth}
    \captionsetup{width=0.75\textwidth}
    \caption{Estimated values ($\beta_l$), standard deviation 
    ($\sigma_{\beta}$), $p$-values and significance (SIG) for 
    logistic regression model with all features.}
    \label{tab-dc:logreg_sum}
    \centering
      \begin{tabular}{|c|r|r|r|r|}
      \hline
      Feature & \multicolumn{1}{c|}{$\beta_l$} & \multicolumn{1}{c|}{$\sigma_{\beta}$} & \multicolumn{1}{c|}{$p$-value} & \multicolumn{1}{c|}{SIG}\\
      \hline
      Intercept  & $-3.30$ & $0.52$ & $2.2e-11$ & ***\\
      \hline
      $\overline{o}_{b,j=1}$  & $22.87$ & $3.32$ & $5.5e-12$ & ***\\
      \hline
      $u_{b,j=2}$  & $-9.51$ & $3.45$ & $0.01$ & **\\
      \hline
      \end{tabular}
    \end{minipage} 
  \end{adjustbox}
\end{table}

\para{Discriminating Features.} 
We perform a model 
selection procedure, based on AIC, to select the most 
discriminating features for our classification model and 
starting from those listed in Table \ref{tab-bd:aggmeasurement}. 
We observe that the feature
$\overline{o}_{b,j=1}=o_{b,j=1}-u_{b,j=1}$ is selected
with high significance. The intuition behind this is that the URLs
which usually come first in bursts are more likely belonging to the
representative class. Thus, $o_{b,j=1}$ is a good distinguishing
feature between representative domains \textit{(SEARCH ENGINE, WEB
  PORTAL)} and non-representative domains \textit{(ADS, CDN)} (as
shown in Figure \ref{fig-bd:measurement}). Solely using this
feature will misclassify domains from \textit{STATIC CONTENT} class as
representative (as these are also likely to come first in burst). This
class includes many CSS HTML pages and static images on 
web-pages. However, the exceptions such as those from \textit{STATIC CONTENT}
class have a high probability of being alone in their bursts, as shown
by $u_{b,j=1}$ in Figure \ref{fig-bd:measurement}. Thus, the
feature $\overline{o}_{b,j=1}$ is able to distinguish between most
representative and non-representative domains.

Note that the domains in the \textit{SEARCH ENGINE} class 
have a unique characteristic, i.e. they show high
values in both $u_{b,j=1}$ and $o_{b,j=1}$ features. 
However, the differenced $\overline{o}_{b,j=1}$ can still act as a
discriminator in selecting representative domains.

The corresponding estimated coefficients
$\beta_l$ are shown in Table \ref{tab-dc:logreg_sum} along with standard
deviation and $p$-values, indicating all significant coefficients. 
As explained above, domains that have high rankings among others, i.e. 
do not appear alone in their bursts, are more likely to be representative domains.
From the estimated value $u_{b,j=2}$, we can also see that 
domains appearing in small bursts of few unique records have smaller chance 
of becoming representative domains.

\subsection{Trade-off Between Classification Metrics}
\begin{figure}[tb]
\begin{adjustbox}{center}
\begin{minipage}[t]{0.65\linewidth}
\centering
\includegraphics[scale=0.22]{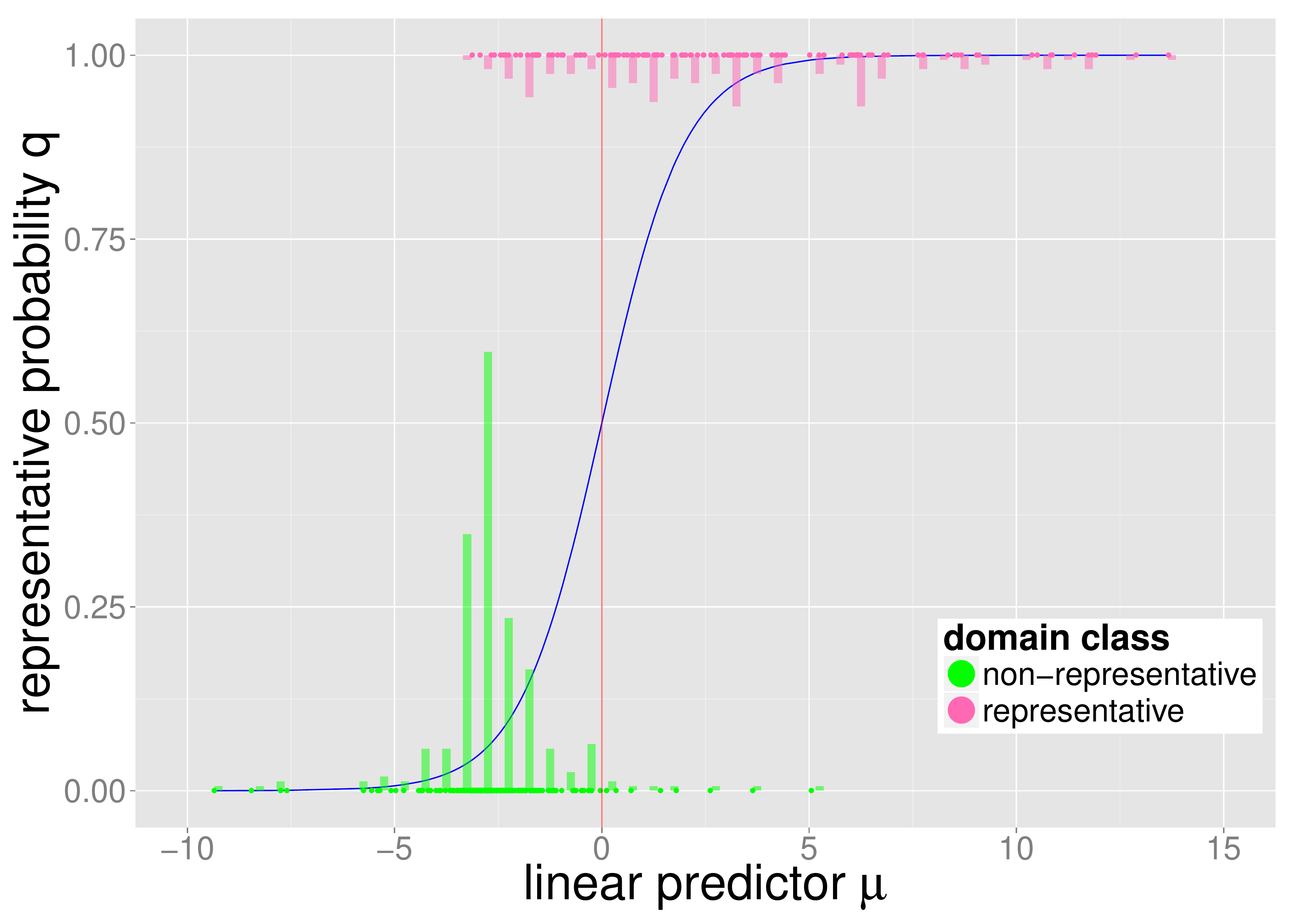}
\vspace{-0.1in}
\caption{Logistic curve of the predictor and the histograms of 
labelled observations.}
\label{fig-dc:logistic_curve}
\end{minipage}
\begin{minipage}[t]{0.65\linewidth}
\centering
\includegraphics[scale=0.22]{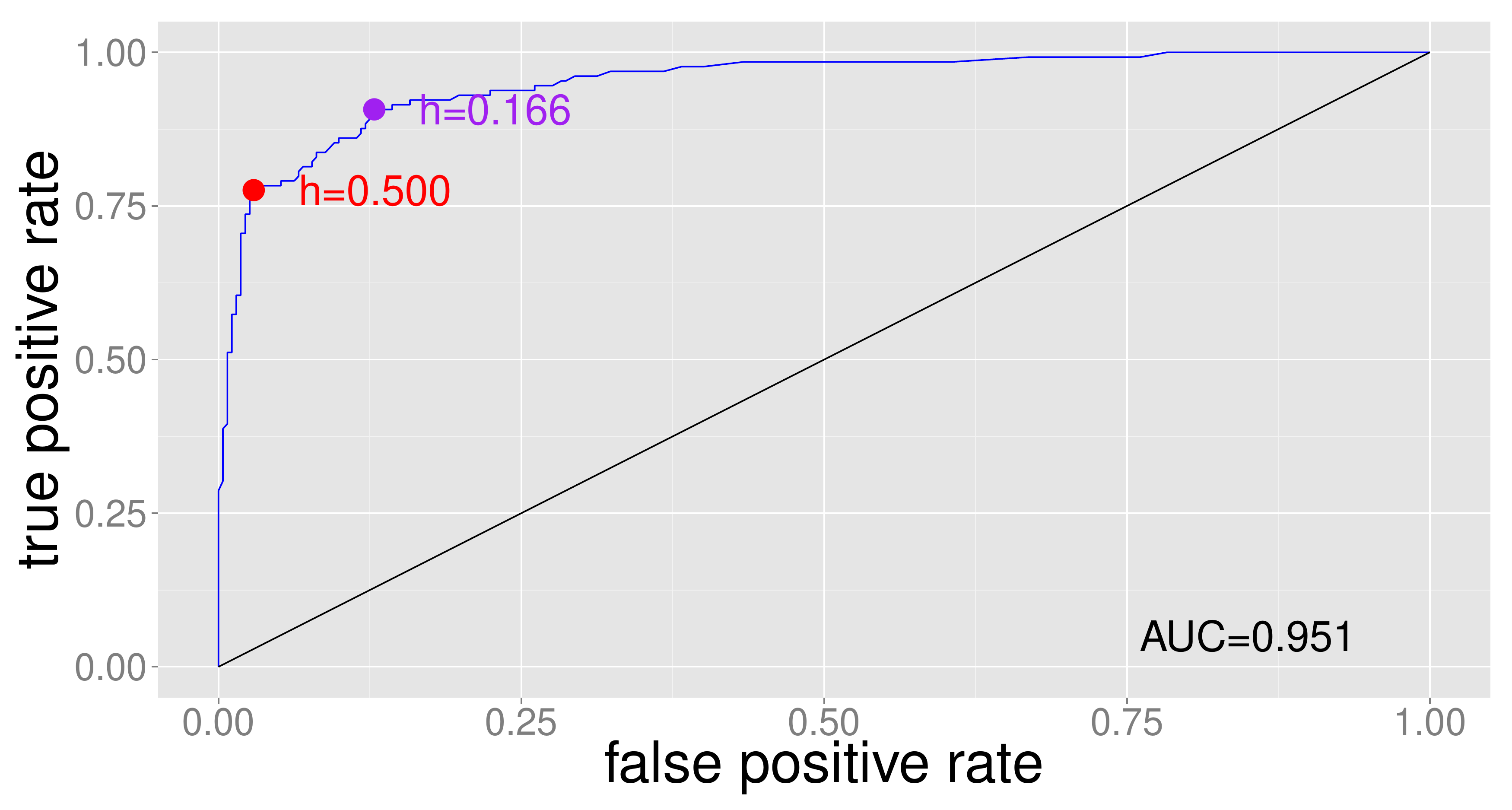}
\vspace{-0.1in}
\caption{Receiver operating characteristic curve.}
\label{fig-dc:roc_curve}
\end{minipage}
\end{adjustbox}
\end{figure}

The relation between the linear predictor $\mu_i$ and representative
probability $q_i$ is plotted in Figure \ref{fig-dc:logistic_curve}, together
with the binary labelled observations and histogram of each domain
class. The red vertical line represents the decision boundary such
that all URL with $q_i\geq h=0.5$ are put into representative class
and the other are in the non-representative class. Hence, the ratio
between the points on the left and right of the red line at row $q=0$
corresponds to the ratio between true negative-ness and false
positive-ness. Similarly, the ratio between true positive-ness and
false negative-ness is at row $q=1$.

%

In Figure \ref{fig-dc:logistic_curve}, we use the boundary value $h=0.5$,
corresponding to the case of minimizing the number of
misclassification cases to separate between representative and
non-representative records. However, from a decision theory
point-of-view, as the false-positive and false-negative penalties are
usually different, we may want to optimize the boundary value 
by customizing these penalties based on application specific requirements. 
For instance, let's consider an application that 
generates users's profiles. These kind of applications may want to put a
higher penalty for false negatives (representative URLs incorrectly
classified as non-representatives) than for false positives
(non-representative URLs correctly classified as representative). This
is because when determining the activity of a burst, there is an
opportunity to prune out the noise (non-representative URLs) further,
while the representative URLs lost in the process are unlikely to be
re-inserted later on. Thus, these applications should be calibrated to 
improve sensitivity, i.e. the ratio of correctly
classified representative URLs to the total number of representative
URLs. 

Because the penalties are
problem-specific and not obvious in many contexts, we show the
trade-off between true positive rate (complement of false negative
rate) and 
false positive rate with the receiver operating characteristic curve
(ROC) in Figure \ref{fig-dc:roc_curve}. The figure also illustrates another
optimal boundary point in purple for the case of minimizing the sum of
false positive and false negative rates. The high value of area under
the curve (AUC),~\cite{Fawcett2006}, again confirms the 
good performance of our classifier.

Table \ref{tab-dc:type12} provides further statistics on the trade-off
between false positives and true negatives for two different values of
the boundary, that were shown in Figure \ref{fig-dc:roc_curve}. These
trade-offs are characterized in terms of precision, negative
predictive value, sensitivity, specificity and accuracy. We note that
while the value of $h=0.5$ results in higher precision and accuracy,
the $h=0.166$ results in better sensitivity. Thus, for applications
with more emphasis on accuracy, we may choose $h=0.5$, while for
applications where sensitivity is crucial, we may select $h=0.166$.

\begin{table}[tb]
  \centering
  \begin{adjustbox}{center}
    \begin{minipage}{.55\linewidth}
    \caption{Trade-off between accuracy, sensitivity, precision and specificity.}
    \label{tab-dc:type12}
    \centering
      \begin{tabular}{|c|r|r|}
      \hline
      $h$ & $0.500$ & $0.166$ \\
      \hline
      Precision & $0.93$ & $0.77$ \\
      \hline
      Negative Predictive Value & $0.90$ & $0.95$ \\
      \hline
      Sensitivity & $0.78$ & $0.91$ \\
      \hline
      Specificity & $0.97$ & $0.87$ \\
      \hline
      Accuracy & $0.91$ & $0.88$ \\
      \hline
      \end{tabular}
    \end{minipage}%
    \begin{minipage}{.55\linewidth}
    \caption{AIC and classification accuracy.}
    \label{tab-dc:accuracy_BM}
    \centering
      \begin{tabular}{|c|r|r|r|r|r|r|r|}
      \hline
      & \multicolumn{1}{c|}{AIC} & \multicolumn{1}{c|}{BIC} & \multicolumn{5}{c|}{Classification accuracy}\\
      \hline
      $\mathcal{C_B}$  & $112.80$ & $122.69$ & $90.0$ & $88.3$ & $90.2$ & $89.2$ & $90.5$ \\
      \hline
      $\mathcal{C_R}$ & $242.82$ & $252.72$ & $69.7$ & $72.9$ & $71.0$ & $70.8$ & $71.4$ \\
      \hline
      \end{tabular}
    \end{minipage} 
  \end{adjustbox}
\end{table}

%
%

\para{Accuracy.} The usage of burst-level features, and in
particular $\overline{o}_{b,j=1}$, results in significant improvement
in the accuracy of the classification model. As shown in
Table \ref{tab-dc:accuracy_BM}, the accuracy improves from around 70\% to
around 90\%; AIC value drops from 242.82 without burst-level
features to just 112.8 with these features; and finally, the BIC drops
from 252.72 to 122.69.

Altogether, these results show that it is possible to achieve a 90\%
accuracy in segregating representative URLs from non-representative
URLs by using burst-level features on truncated URL web-traces. In
other words, the burst decomposition and the extraction of specific
features from the bursts overcome the information lost due to URL
truncation. Note that this accuracy is in terms of the number of URLs correctly
classified as being representative or non-representative. Popular URLs
are more likely to be correctly classified by our methodology and
thus, the accuracy in terms of number of records, download size (e.g.,
to answer questions like how much download is generated corresponding to each
activity type) or number of bursts (identifying the activity for each
burst) is likely to be significantly higher.

\section{Related Work}
\label{sec:relatedwork}

The past research related to identifying URLs representing user activities falls into the following categories:
\begin{enumerate}
\item
Providing an activity description at a very coarse level (e.g., peer-to-peer
networking, HTTP browsing, chatting etc.) by filtering out URLs based on connection 
port number, packet payload, statistical traffic patterns etc. \cite{Nguyen2008,Zhang2013}, primarily for the purpose of network traffic analysis and traffic classification (e.g. for CDN).
\item
Manual blacklisting of URLS to filter out spam, adult content or advertisements.
\item
Filter out the
unintentional traffic by relying on URL suffixes (e.g., .mp3, .js etc.), URL
header patterns, HTTP referrers and HTTP blacklists.
\end{enumerate}

The first category of coarse-grained application-type classification is different from our representative URL identification problem as our goal is to segregate URLs at the HTTP domain-level (HTTP browsing itself is just one class for the coarse-grain classification). From the behavioral analysis' perspective, traffic classification \cite{Nguyen2008,Zhang2013} cannot provide a medium-grained insight into user web activities such as reading, shopping, researching, etc. 

The second category has obvious scalability limitations. Given that many new web-sites appear every day in different languages and different countries that may have very local characteristics, it is very expensive to manually maintain the blacklists. Furthermore, the existing blacklists (see~\cite{manualBlacklist} for a list of many manual blacklists) are for specific purposes (such as spam, adult content, advertisements) and do not contain all non-representative URLs (such as images and videos associated with the main content). In fact, it is difficult to even filter out all unintentional traffic using only these blacklists.

The third category critically relies on full HTTP web-traces. For some papers (e.g.,~\cite{Zhang2011}) in this category, the setting even allows to take a peek into a user's full network traffic (including a deep packet inspection of the content). The full HTTP URLs can reveal highly sensitive user information and their usage raises serious privacy concerns. For instance, Song et al. \cite{Song2013}  highlighted a practical privacy attack that exploits seemingly-anonymous recorded information of shortened URL service such as  HTTP referrer URLs, countries, browsers, platform, etc to infer the clicking pattern of a specific user. 
In contrast, our focus is on inferring medium-grained user behavior analysis from minimal traffic traces (truncated-URLs) and on techniques that will allow us to offset the accuracy loss due to URL truncation. 

Furthermore, our work deals with a large, diverse, but noisy traffic
trace from a network-side. This allows us to perform a detailed study 
that is not limited to a few domains or restricted to a few volunteer
users. This is in contrast with many publications on behavior analysis that
deal with data from users or service-providers.




\textbf{Privacy preserving user profiling.} There has been considerable work in recent years on privacy preserving personalization. Herein, we list a few approaches:

Bilenko and Richardson~\cite{BR11} also consider the problem of user profiling while mitigating the privacy concerns. However, their approach is based on storing the sensitive information on the client side, in the form of cookies or browser local storage. The storage of this sensitive information still leaves a user vulnerable to privacy violations. On the other hand, our user profiling does not require the sensitive information to be stored at all. Similar client-side approaches in the context of personalized search (e.g.,~\cite{XWZC07}) and online advertisements (e.g.,~\cite{TNBNB10}) have also been studied.

Nandi et al.~\cite{NAB11} take an alternative approach to privacy preserving personalization. They replace the user traces by traces of group of similar users. However, this requires user segmentation, which in turn, requires significant historical data. Also, this results in an aggregate level personalization and not an individual level personalization.

Also, there are some theoretical solutions based on k-anonymity~\cite{Swe02} and l-diversity~\cite{MKGV07} for preserving privacy. However, it is not clear if they can be useful for profiling personalized time-series data. There are also some approaches (e.g.~\cite{LiSPMS07}) that add random or correlated noise to the data to preserve the privacy. However, such approaches also introduce more noise in the user profiles.


\textbf{Burst Detection.} Kleinberg \cite{Kleinberg2002} proposed a discrete state space model
as a burst detection algorithm, with applications in email
streams. However, the focus of this solution is the varying
exponential distribution's rate, modelled by the hidden state. The
rate characterizes the email arrival of a temporal local period but
does not provide a clear distinction between within-burst and
out-of-burst records. For example, even when the rate goes down to the
smallest value, the positive skewness of exponential distribution
still favours small inter-arrival time samples. Such an approach is
unlikely to work for our
problem of segregating two inter-arrival time classes. 
 
Crovella and Bestavros \cite{Crovella1997} and Wang et
al. \cite{Wang2002} looked at the related problem of counting process
modeling of download size, open connection, disk operation request,
etc. However, these papers primarily focus on the self-similarity
property of the time series and do not provide a clear distinction and
separation of different HTTP record types.  

Karagiannis et al. \cite{Karagiannis2004} showed that the accuracy of
exponential distribution varies with different backbone packet
traces. In general, exponential distribution has nice mathematical
features such as memoryless-ness and closed-form solutions of
sum-concat-minimum operators. However, its light tailed property may
not be a good match to some datasets. In a more general context of
human dynamics, Barabasi \cite{Barabasi2005} discussed the bursty
nature of human actions and argued that heavy tailed distribution
pareto is a better match than exponential distribution with email data
analysis. However, both of these papers analyze aggregated datasets,
and not the per-user dataset which is much more dynamic.

\section{Discussion}
\label{sec:conclusion}
We have proposed a novel methodology to identify URLs representing
user activities from a truncated URL web-trace. Our statistical
methodology offsets the loss in accuracy due to URL truncation by
considering additional features derived from the burst measurements. To enable
the computation of burst-level features, we propose a novel
technique for burst decomposition.

Once the set of representative URLs is identified, one can compare the
(live) streaming web-traces of users to infer medium grained
activities in real-time and offer personalized services. Burst
decomposition can play a critical role in this part as well. Once the
user-adaptive thresholds are identified, our burst decomposition
algorithm can be used to decompose the streaming trace into bursts and a
unique URL representing the activity in that burst can be identified
using the identified set of representative URLs.

We consider that our methodology can be very useful for providing
personalized services, while being considerate about more sensitive
user privacy data. For instance, state-of-the-art techniques to predict click-through-rate (CTR) rely on behavioral targeting of fine-grained user data~\cite{CPC09}, such as advertisement clicks, web-page clicks, page views and search query data. A medium-grained user profiling, such as the one created by our technique, can be used to provide good CTR predictions while preserving privacy considerations. Similarly, user segmentation based on behavioral targeted advertisement (e.g.,~\cite{YLWZJC09}) can also benefit from our medium-grained profiling. Our profiling can also be employed to re-rank the search results for a more personalized experience (similar to the approach in ~\cite{XWZC07}). More generally, we hope that our work will lead to
deeper studies on the usage of truncated URL traces, as a means to
striking the fine balance between personalized services and user privacy.



\bibliographystyle{splncs03}
\bibliography{biblo}
\newpage
\appendix
\section{Modeling Inter-arrival Time Distribution}
\label{sec:model-inter-arrival}
We consider 
the models of exponential and Pareto distribution
functions to fit the inter-arrival time samples of
all users. 
Specifically, we use the following distribution functions:
\begin{itemize}
\item
Exponential Distribution (EXP):
\begin{align}
\label{eqn-bd:pexp}
p_{EXP}(x) &= f_{\lambda}(x) = \lambda \exp(-\lambda x) ,
\end{align}
where $\lambda$ is the inter-arrival rate of all 
records.
\item
Pareto Distribution (PRT):
\begin{align}
\label{eqn-bd:prt}
p_{PRT}(x) &= g_{\alpha}(x) = \alpha x_m^{\alpha} x^{-\alpha-1} ,
\end{align}
where $x_m$ is the minimum value of the Pareto distribution and $\alpha$ is the shape
parameter. 
\end{itemize}
In addition, we use two mixture models. The intuition behind the
mixture approximation is that the two models will respectively capture
the high arrival rate for the within-burst records, and the slow
arrival rate for the out-of-burst records.
\begin{itemize}
\item
Mixture of two Exponential Distributions (EXP2):
\begin{align}
\label{eqn-bd:pexp2}
p_{EXP2}(x) = w f_{\lambda_1}(x) + (1-w) f_{\lambda_2}(x) ,
\end{align}
where $\lambda_1$ and $\lambda_2$ are the corresponding 
inter-arrival rate parameters for within-burst and out-of-burst
records with $\lambda_1>\lambda_2$; $w$ is the proportion of within-burst records.
\item
Mixture of two Pareto Distributions (PRT2):
\begin{align}
\label{eqn-bd:prt2}
p_{PRT2}(x) = w g_{\alpha_1}(x) + (1-w) g_{\alpha_2}(x) ,
\end{align}
where $\alpha_1$ and $\alpha_2$ are the corresponding 
Pareto shapes of within-burst and out-of-burst records; 
$w$ is the proportion of within-burst
records.
\end{itemize}
We observed that the exponential component matches with the within-burst 
records while the heavy tailed Pareto component is better for 
out-of-burst records. Hence, we use another function form:
\begin{itemize}
\item
Concatenation of Exponential and Pareto Distributions (EXP\_PRT):
\begin{align}
\label{eqn-bd:exp_prt}
p_{EXP\_PRT}(x) &= z[\exp(-\lambda x)I(x \leq d) + c x^{-\alpha-1} I(x > d)] ,
\end{align}
where $I(\cdot)$ is the indicator function, $z$ is the density 
normalization constant and the constant $c$ is to make 
the function continuous.
\end{itemize}
These five distributions are fitted with per-user inter-arrival records and all the
parameters $\lambda$, $\alpha$, $w$ and $d$ for each user are
estimated by Maximum Likelihood Estimation (MLE) method. 

Note that our goal is not just to 
model the inter-arrival density function, but to model it in a
parameterized way that provides an intuition for 
identifying the separation threshold for
burst decomposition. Kernel density approximation may be well-matched
for the target density but cannot be used as it does not imply any
meaning of record types.  

The mixture models of the standard exponential and Pareto functions
portray such an intuition. For example, in the method EXP2, the
exponential component with high arrival rate is supposed to contain
all within-burst records; So, its $99\%$ quantile can be used as a
threshold to separate the records types. Or in the method EXP\_PRT,
the location parameter $d$ can be used as the threshold as it is the
boundary of the exponential component for within-burst records and the
heavy-tail Pareto distribution for out-of-burst records.

\subsection{Model Selection and Evaluation}
We consider the inter-arrival time distribution of various users to
determine (i) whether or not our basic intuition of 
a user's data traffic as essentially consisting of a series of burst
is true, (ii) which of the considered functions best fits the inter-arrival time
distribution of users and (iii) how well does the best fitting
function matches the inter-arrival time distribution of users.  

We observed that for most users, there are two well-separated 
components in the inter-arrival time distributions. Furthermore, for 
most users, the exponential function tightly
matches the within-burst component and the tail of the
Pareto function closely matches the portion of human activity 
with long delay, corresponding
with the out-of-burst records. In addition, the positive skewness of both
exponential and Pareto distributions implies the existence of compacted
within-burst records, which have small inter-arrival time. All of
these observations confirm our basic intuition that a user's
data-traffic primarily consists of a number of contiguous 
bursts of URL records. Furthermore, the two components of 
within-burst and out-of-burst records in the inter-arrival time 
distribution are usually separable. 

\para{Determining the Best-fit Model.} Next, we consider the modeling of
inter-arrival time distributions. We compare the fitness of all the studied 
distributions by using Akaike information criterion (AIC) \cite{Akaike1974}, based on
the Kullbeck-Leiber discrepancy between the true unknown model and the
best estimated model of the assumed family and Bayesian information
criterion (BIC) \cite{Schwarz1978}, based on the Laplace approximation of the marginal
likelihood. The EXP\_PRT model, presented in 
Equation 8
, shows the best AIC and BIC values, i.e. the
smallest discrepancy between the data and the model, which we have
highlighted in bold in Table \ref{tab-bd:aicbic}. Thus, we conclude that
the EXP\_PRT model best fits the target inter-arrival time
distribution, with its exponential function capturing the within-burst
component and its Pareto function matching the out-of-burst component.

\begin{table}[t]
\caption{Evaluation of AIC/BIC ($\times 1e^6$) of different density approximations.}
\centering
\begin{tabular}{c|r|r|r|r|r|}
\cline{2-6}
& \multicolumn{1}{c|}{EXP} & \multicolumn{1}{c|}{EXP2} & \multicolumn{1}{c|}{PRT} & \multicolumn{1}{c|}{PRT2}  & \multicolumn{1}{c|}{EXP\_PRT}\\
\hline
\multicolumn{1}{|c|}{AIC} & $291.14$ & $179.77$ & $118.20$ & $118.21$ & $\bf{59.59}$ \\
\hline
\multicolumn{1}{|c|}{BIC} & $291.17$ & $179.86$ & $118.23$ & $118.30$ & $\bf{59.68}$ \\
\hline
\end{tabular}
\label{tab-bd:aicbic}
\end{table}


\begin{figure}[!ht]
\centering
\begin{adjustbox}{center}
  \begin{subfigure}[t]{0.42\textwidth}
    \centering
    \includegraphics[scale=0.22]{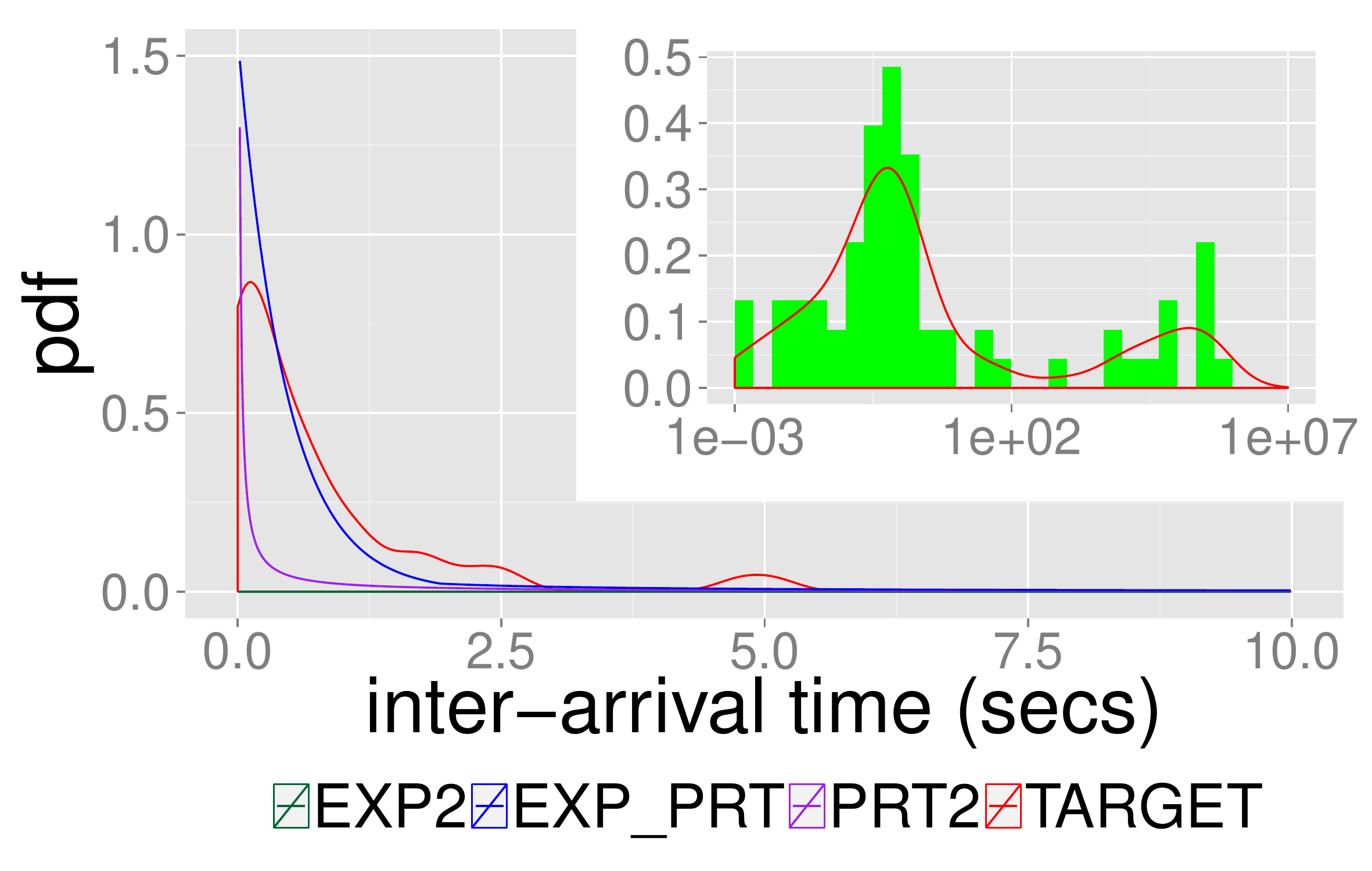}
    \subcaption{Low activity user: $71$ records.}
    \label{fig-bd:iatden_A}
  \end{subfigure}
  \begin{subfigure}[t]{0.42\textwidth}
    \centering
    \includegraphics[scale=0.22]{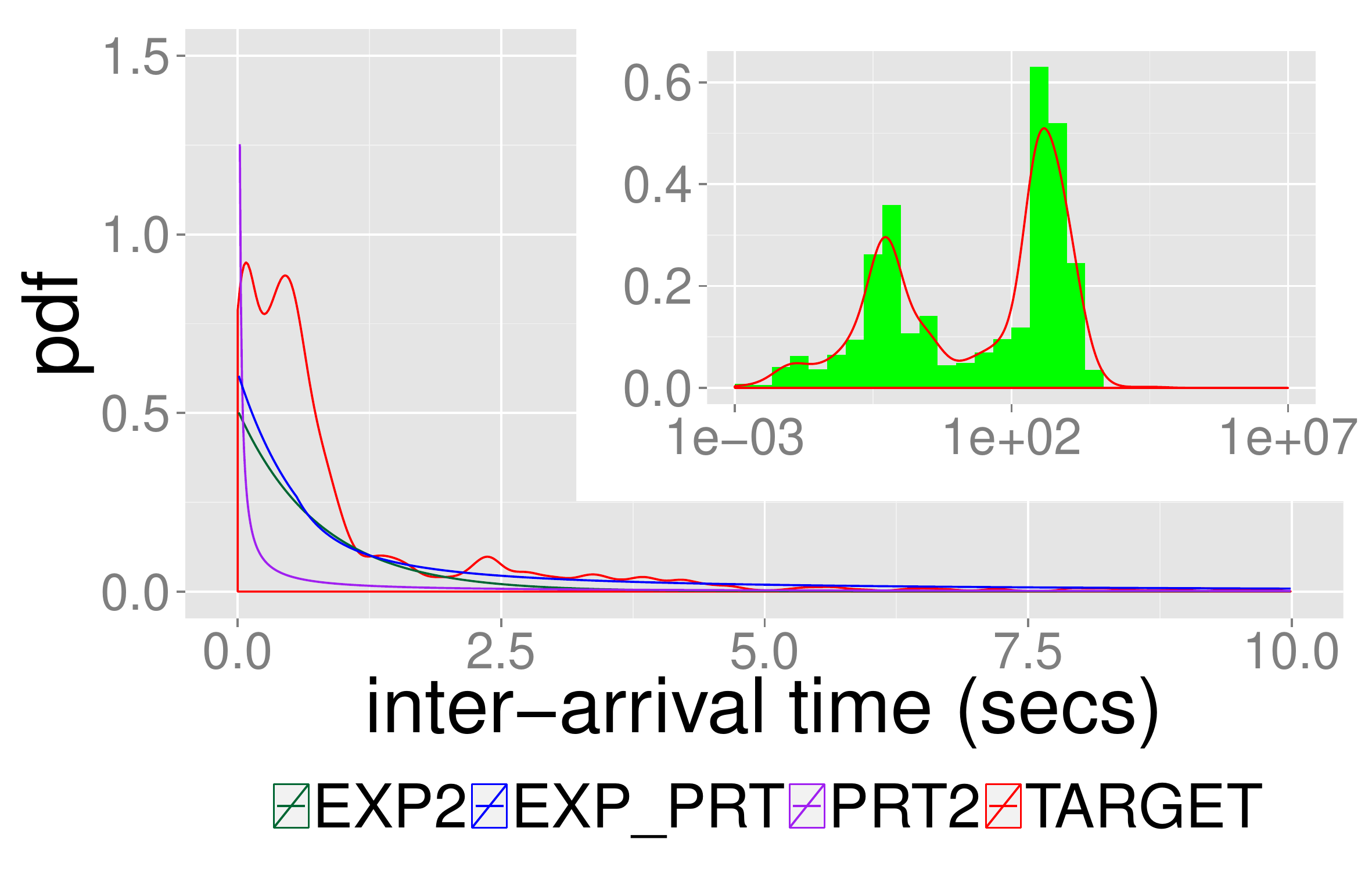}
    \subcaption{Medium activity user: $6764$ records.}
    \label{fig-bd:iatden_C}
  \end{subfigure}
  \begin{subfigure}[t]{0.42\textwidth}
    \centering
    \includegraphics[scale=0.22]{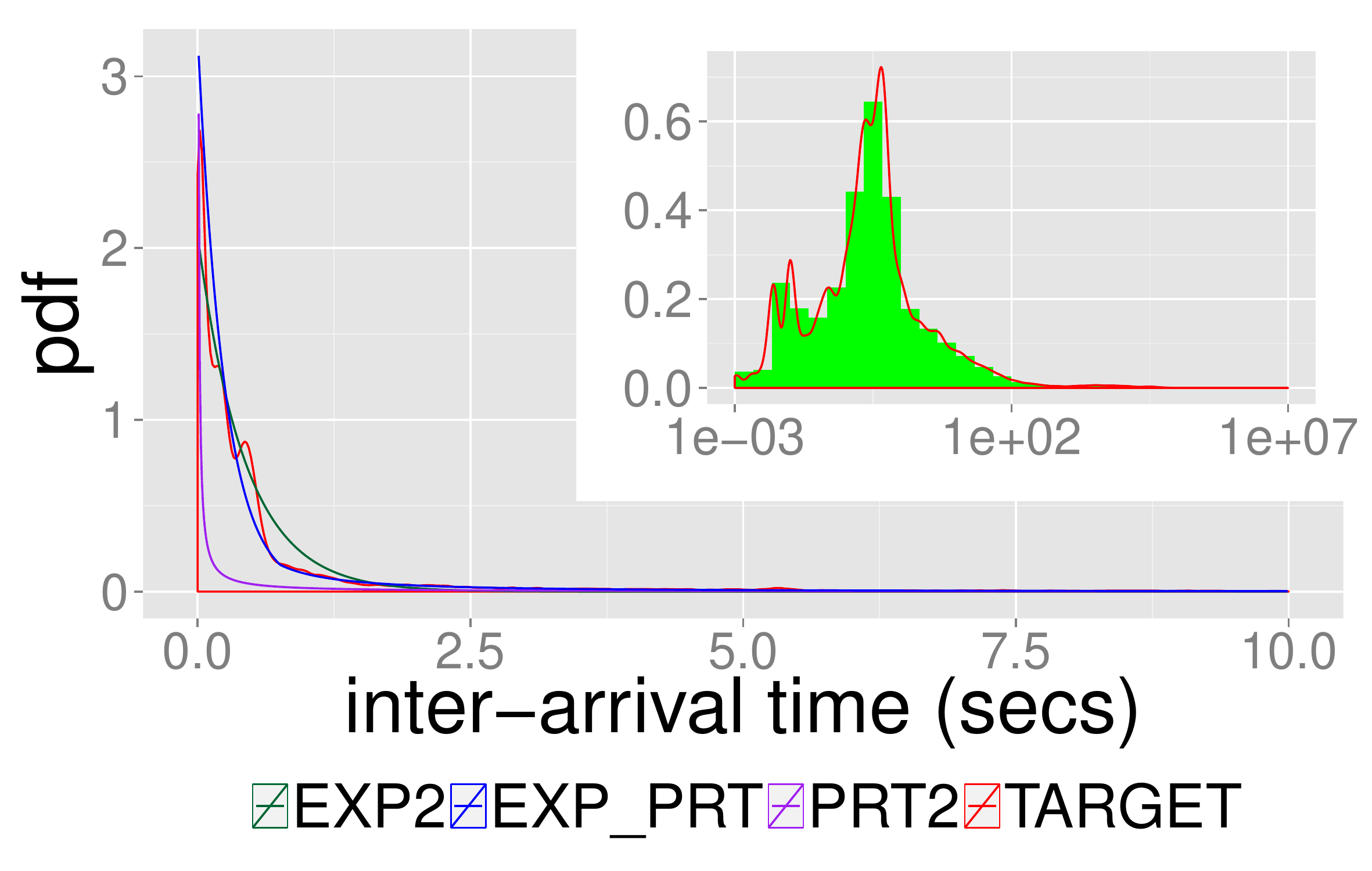}
    \subcaption{High activity user: $35116$ records.}
    \label{fig-bd:iatden_B}
  \end{subfigure}
\end{adjustbox}
\caption{Density approximation of inter-arrival time 
  with inner log scale density plots (target curve is plotted in
  red).}
\label{fig-bd:iatden}
\end{figure}

\textbf{Fitting Error of the EXP\_PRT Model Across All Users.} We
investigate whether the EXP\_PRT model is able to capture the
diversity across users with high level of accuracy. As we are going to
show next, this is not always true and so, a robust technique to
separate within-burst and out-of-burst records
that is independent of user-based distribution shape is needed. 

Figure \ref{fig-bd:iatden} shows
the distribution fitting of EXP2, PRT2 and
EXP\_PRT functions for three different users. We find that out of the
three users considered in this figure, the
EXP\_PRT function fits the target inter-arrival time distribution of 
two users (Figures \ref{fig-bd:iatden_A} and \ref{fig-bd:iatden_B}) very
well, but it is a poor match for the third user (Figure \ref{fig-bd:iatden_C}). We
quantify the density approximation error by the following measure:
\begin{align}
\label{eqn-bd:errmeasurement}
S_b &= \int_x |p(x)-q(x)| dx,
\end{align}
Here, $p(x)$ is the target kernel density as interpolated (from the inter-arrival time
distribution of the user) by a kernel
density approximation function. The function $q(x)$ is the one 
that we use to fit our target distribution. The error measure is
calculated by numerical integration.

The value of our error metric ranges from $0$ to $2$, where $0$ corresponds to a
perfect fit. The closer the error statistic is to $1.0$, the poorer is the fit,
and a value of greater than $1.0$ reflects an extremely poor fit. The
error measurements of the method EXP\_PRT for 
Figures \ref{fig-bd:iatden_A}, \ref{fig-bd:iatden_C} and \ref{fig-bd:iatden_B} are $0.46$,
$0.84$ and $0.26$ respectively. 

\begin{figure}[!ht]
\centering
\includegraphics[scale=0.22]{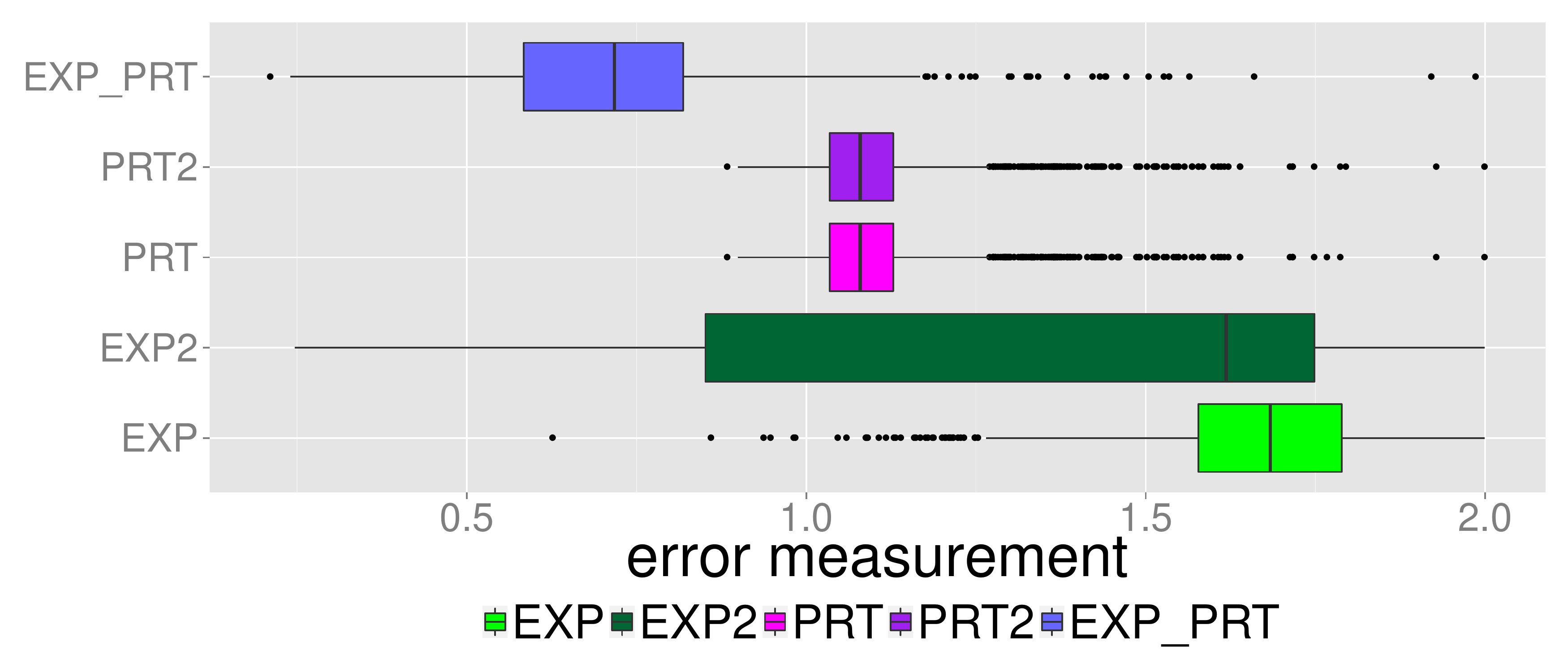}
\caption{Density approximation errors of different methods.}
\label{fig-bd:errmeasurement}
\end{figure}

Figure \ref{fig-bd:errmeasurement} presents the variation in the value of
our error metric 
for various fitting functions, over all users. Again, we observe that
the EXP\_PRT function results in the lowest error (even with this
error measure). However, this is still a poor fit. The median user
has an error of $0.72$, which means that about $50\%$ of the users
have approximation errors $S_{b,EXP\_PRT}$ larger than $0.72$. We
consider that this is too high an error to be useful for developing an
analytics system using it. 
More specifically, the thresholds based on such a
poorly fitting function are unlikely to result in a good burst decomposition.

The above analysis suggests that these density functions are not
flexible enough to accommodate highly varied and personalized
inter-arrival time of different users. Thus, there is a need for a 
robust technique to separate within-burst and out-of-burst records
that is independent of the personalized distribution shape of a user.

\end{document}